\documentclass[twocolumn]{aastex63}
\usepackage{natbib}
\usepackage{graphicx}
\usepackage{float}
\usepackage{latexsym}
\usepackage{longtable}
\usepackage{lineno}

\begin{document}

\title{Confirming Near- to Mid-IR Photometrically-Identified Obscured AGNs in the JWST era}

\author[0000-0003-2303-6519]{George H. Rieke}
\affiliation{Steward Observatory, University of Arizona,
933 North Cherry Avenue, Tucson, AZ 85721, USA}

\author[0000-0001-6561-9443]{Yang Sun}
\affiliation{Steward Observatory, University of Arizona, 933 North Cherry Avenue, Tucson, AZ 85721, USA}

\author[0000-0002-6221-1829]{Jianwei Lyu}
\affiliation{Steward Observatory, University of Arizona,
933 North Cherry Avenue, Tucson, AZ 85721, USA}

\author[0000-0001-9262-9997]{Christopher N. A. Willmer}
\affiliation{Steward Observatory, University of Arizona, 933 North Cherry Avenue, Tucson, AZ 85721, USA}

\author[0000-0003-3307-7525]{Yongda Zhu}
\affiliation{Steward Observatory, University of Arizona, 933 North Cherry Avenue, Tucson, AZ 85721, USA}

 \author[0000-0002-5104-8245]{Pierluigi Rinaldi}
\affiliation{Steward Observatory, University of Arizona, 933 North Cherry Avenue, Tucson, AZ 85721, USA}

\author[0000-0002-9720-3255]{Meredith A. Stone}
\affiliation{Steward Observatory, University of Arizona, 933 North Cherry Avenue, Tucson, AZ 85721, USA}

\author[0000-0003-4565-8239]{Kevin N. Hainline}
\affiliation{Steward Observatory, University of Arizona, 933 North Cherry Avenue, Tucson, AZ 85721, USA}

\author[0000-0003-4528-5639]{Pablo G. P\'{e}rez-Gonz\'{a}lez}
\affiliation{Centro de Astrobiolog\'{i}a (CAB), CSIC–INTA, Ctra. de Ajalvir km 4, Torrej\'{o}n de Ardoz, E-28850, Madrid, Spain}



\begin{abstract}

We evaluate the underlying assumptions for the identification of Active Galactic Nuclei (AGNs) through near- and mid-infrared photometry and spectral energy distribution (SED) fitting out to z $\sim$ 3. For massive galaxies, $log(M) \ge 9.5$, our high resolution spectra of the rest optical range generally confirm the results of SED fitting, which relies primarily on excesses above the stellar emission between 1 and 6 $\mu$m to identify AGN. However, the method is undermined if the redshift used for the SED fitting is incorrect. Low mass galaxies, $log(M) < 9.5$, can contain relatively warm dust that emits in the 4 - 6 $\mu$m range. We show that the potential contamination of AGN samples by purely star forming low-mass galaxies  can be avoided by the use of the infrared properties of Haro 11 as a limiting star-forming SED template. However, relatively few star forming galaxies emit as strongly in the 3--6 $\mu$m range as this template, so this could result in missing some obscured AGNs to avoid a minor contamination. Including the behavior of the galaxies at rest $\lambda \sim$ 13.5 $\mu$m can mitigate this problem and yield more complete  samples of {\it bona fide} AGN. JWST/MIRI supports this approach out to z $\lesssim$ 0.6.

\end{abstract}

\section{Introduction}

Two recent review articles, \citet{Hickox2018, Lyu2022}, have emphasized that mid-infrared observations are critical to find complete samples of heavily dust-obscured AGNs. They describe  large-scale searches using data from IRAS, ISO, Spitzer, and WISE. With the advent of JWST, such searches have been extended to higher redshifts, much lower luminosities, and more sophisticated analyses to take advantage of the sensitivity and many mid-infrared bands provided by the Mid-Infrared Instrument (MIRI) \citep{Yang2023, Kirkpatrick2023,  Lyu2024, Li2024, Chien2024}.

These JWST surveys push the AGN mid-infrared selection into previously unexplored realms.  The assumption behind the new surveys is that the infrared emission powered by star formation should exhibit strong mid-infrared Polycyclic 
Aromatic Hydrocarbon (PAH) features. PAH features do not emit strongly at wavelengths short of 6 $\mu$m (excepting a moderately strong feature at 3.3 $\mu$m) and even obscured AGNs do emit in this range. This difference is the foundation of traditional obscured AGN identification via mid-infrared SED fitting using templates including the PAH bands. This approach should be robust so long as accurate redshifts are available and any infrared excess powered by young stars is PAH-dominated and not so strong that it overwhelms the AGN contribution. 

As an example, \citet{Lyu2024} conducted a comprehensive search for AGN in SMILES. This is a MIRI survey of a central region in the GOODS-S field and coincides with the JADES/NIRCam survey from 0.9 to 5 $\mu$m (plus deep Chandra, HST, ALMA, and JVLA data); see \citet{Rieke2024} for a description. 
For M$_{host} > 5 \times 10^9$ M$_\odot$, a PAH - based template is appropriate to fit stellar-powered infrared emission \citep{Shivaei2024}. \citet{Lyu2024} used such a  template to characterize  infrared emission from host galaxies with M$_{host} > 5 \times 10^9$ M$_\odot$ and found 111 AGN candidates. This was facilitated because SMILES includes all the MIRI bands from 5.6 through 25.5 $\mu$m with integration times  selected to identify PAH features and support this type of study \citep{Rieke2024}.

SED fitting with MIRI photometry also reveals a large number of AGN candidates in lower-mass galaxies: \citet{Lyu2024} identified 86 candidates ($\sim$ 2 arcmin$^{-2}$), whereas \citet{Yang2023} found even more.  However, confirming their nature from photometry alone is challenging, since local low-mass, dwarf galaxies have weak PAH emission,  due to destruction by the hard interstellar radiation fields \citep[e.g.,][]{Hunt2010} and also can have a warm non-PAH infrared component \citep[e.g.,][]{Wu2006, Remy2015} whose emission extends short of 6 $\mu$m. This warm emission arises because of their hard interstellar radiation fields, the porous structure of their ISMs \citep{Madden2016} that allows these fields to penetrate efficiently, and because their  dust contains very small grains, for example from shattering by shocks (see discussion in the review by \citet{Hunter2024}). 
In a study of local star-forming galaxies with M$_* < 3 \times 10^9$ M$_\odot$, \citet{Hainline2016} found the extreme case is defined by a  number of galaxies with significant excesses even in WISE $W1-W2$, i.e., at 4.6 $\mu$m. From these examples, they conclude that star formation can heat dust to temperatures where the emission can be confused with AGN heating. 

Locally, the low-mass, hot-dust galaxies identified by \citet{Hainline2016} appear to be rare. Nonetheless, searches for obscured AGNs need to be aware of the possibility that a larger fraction of low-metallicity galaxies at the relevant redshifts may have sufficient hot dust to violate selection criteria based on PAH-inclusive star forming templates. At the same time, if hot dust examples are indeed rare even at these  redshifts, selection methods need to be tuned so they do not reject a large number of true AGN to avoid a low level of contamination.  

This paper examines the issues around AGN identification in both high-mass and low-mass galaxies. As part of the search for obscured AGN in the SMILES survey  reported by \citet{Lyu2024}, we obtained NIRSpec Micro Shutter Array (MSA) \citep{Jakobsen2022} spectra of a small sample of objects. This sample was designed as a test of whether the source properties are as assumed in the photometric SED fitting. Specifically, we wanted to test whether 
(1)  the star formation rates deduced from the spectra are consistent with those deduced from the SED fitting; and (2) the AGN emission lines are suppressed in the purported obscured sources, consistent with the level of obscuration. The spectra also give us an insight to the behavior of the SED fitting in identifying AGN at redshifts z $>$ 2.5, where the aromatic features move out of the MIRI photometric bands. This paper will discuss these spectra in Section 3. 

The second objective of the paper is an  assessment of the risk that low metallicity galaxies can mimic AGNs in infrared photometry, leading to false identifications such as those discussed by \citet{Hainline2016}. In Section 4, we will  evaluate the behavior of local dwarf galaxies in AGN color selections and the prevalence of hot-dust SEDs that might be confused with AGNs. We suggest  a suitable SED template where hot dust is suspected and  then suggest a method to further refine mid-infrared photometric AGN searches to  moderate redshift, using MIRI photometry. Finally, Section 5 discusses the issue of degeneracy in photometric AGN searches, followed by a concluding summary. 

\section{NIRSpec observations} 

As summarized in Zhu et al. (in prep), we have obtained JWST/NIRSpec MSA spectra of 166 galaxies as part of the SMILES survey (PID 1207, PI: Rieke; \citealt{Alberts2024,Rieke2024}) with the main goal to characterize the spectral properties of various MIRI sources. These galaxies were observed with the G140M/F100LP and G235M/F170LP gratings and blocking filters to cover 0.97-3.07 $\mu$m in the observed frame (0.4-1.3 $\mu$m in the rest frame for a typical redshift in our sample) at a spectral resolution of $R\sim 1000$. The effective exposure time with each disperser/filter combination was 7,000 seconds ($\sim 1.94$ hours). In this work, we focus on a subset of 17 MIRI-selected AGN candidates with $z$=1.24--3.33 so that the [O III], H$\beta$, [N II] and H$\alpha$ can be covered with the instrument setup.   

We processed and calibrated the data using the {\tt JWST Calibration Pipeline} \citep{Bushouse2022}, version 1.14.0, with the CRDS\footnote{Calibration Reference Data System: \url{https://jwst-crds.stsci.edu/}} context {\tt jwst\_1236.pmap}. In brief, the {\tt calwebb\_detector1} stage converted ``uncal.fits'' files into uncalibrated rate images. Next, the {\tt calwebb\_spec2} stage applied flat-field corrections, flux calibration, and local background subtraction, yielding resampled spectra for each nodding position. The final spectral combination and extraction were carried out using {\tt calwebb\_spec3}. Additionally, we applied custom scripts to further reject hot pixels and remove $1/f$ noise. A detailed description of the reduction process, along with reduced spectra and the redshift catalog, will be provided in a forthcoming data release (Y.~Zhu et al., in prep.; see also \citealt{Zhu2025}). The data can be accessed at doi=10.17909/et3f-zd57 and the spectra are displayed in the appendix.

The MSA shutters provide slitlets of size $0.20 \times 0.46''$ in size. We did not apply a slit-loss correction for line ratio measurements in BPT diagrams to avoid any bias due emission lines having different morphologies in the galaxy images used for the correction. For H$\alpha$ luminosity, however, we corrected for slit loss by comparing the flux within the MSA shutter to the total Kron-convolved photometry across multiple NIRCam bands (F115W, F150W, and F210M), and we fit a second-order polynomial to model wavelength-dependent slit loss. 
Emission lines were measured using Gelato \citep{Hviding2022}, from which line parameters (redshift, intensity, widths) were determined. The emission line template includes [OII] $\lambda\lambda$3727\&3729, [OIII] $\lambda\lambda$4959\&5007, [OI] $\lambda\lambda$6302\&6365, [NII] $\lambda\lambda$6549\&6585, [SII] $\lambda\lambda$6718\&6732, and Balmer lines modeled by a single Gaussian component each. 
We used the spectroscopic redshift, based on all the lines well detected in the spectrum, as the input prior.  A total of 17 galaxies that were identified as having  obscured AGN from SED analysis \citep{Lyu2024} were included in  this analysis, selected to illustrate a range of behavior and redshift. Key emission lines and their ratios for these galaxies, along with the method of identifying AGNs and estimates of their metallicities, are summarized in Table~\ref{gallist}.

\begin{deluxetable*}{lcccccccccccc}
\tabletypesize{\footnotesize}
\label{gallist}
\tablecaption{Key lines and ratios} 
\tablewidth{0pt}
\setlength{\tabcolsep}{2.5pt}
\tablehead{
\colhead {JADES} & 
\colhead {z} &
\colhead{SMILES} &
\colhead{AGN } &
\colhead {H$\alpha$\tablenotemark{c}}  &
\colhead {err}  &
\colhead{H$\alpha$/H$\beta$}  &
\colhead {err} &
\colhead {[OIII]\tablenotemark{d}/H$\beta$} &
\colhead {err}  &
\colhead {[NII]\tablenotemark{e}/H$\alpha$} &
\colhead {err}  &
\colhead {Z$_\odot$\tablenotemark{f}} 
 \\
\colhead {ID} &
\colhead {}  &
\colhead{ID\tablenotemark{a}} &
\colhead{Code\tablenotemark{b}} &
\colhead {10$^{-17}$W/m$^2$}  &
\colhead {} &
\colhead {} &
\colhead {} &
\colhead {}  &
\colhead {}  &
\colhead {}  &
\colhead {}  &
\colhead {}  
}
\startdata
 87191\tablenotemark{g} &  2.15 &  540 &  1 & 6.12 & 0.011 &  7.09 &  0.45 & 14.14 &  0.29 &  0.01 &  0.001 & 7.9  \\
113325\tablenotemark{g} &  2.49 & 2573 &  1  & 1.042 & 0.013 &  3.02 &  0.15 &  5.19 &  0.25 & -0.02 &  0.01 & --  \\
194121 &  1.98 &  607  &  1 &  5.03 & 0.022 &  3.65 &  0.10 &  2.75 &  0.08 &  0.15 &  0.01 & 8.3   \\
194473 &  1.61 &  634  &  1,3  & 0.884 & 0.067 & 16.39 &  7.92 &  7.76 &  1.24 &  1.22 &  0.41 & 8.4   \\
196184 &  2.65 &  693  &  1,3  &  3.21 & 0.036 &  8.71 &  2.60 &  2.98 &  0.33 &  0.41 &  0.01  &  8.3  \\
196290 &  2.59 &  701  &  1,2,3,4  &  19.2 & 0.045 & 13.97 &  4.66 & 21.00 &  2.48 &  0.42 &  0.01  &  8.2    \\
201027 &  2.87 &  920  &  1  &  1.49 & 0.023 &   Inf &  7.80 &  --- &  --- &  1.35 &  0.05  & ---\\
202378 &  2.62 &  972  &  1  &  7.26 & 0.026 &  7.86 &  2.05 &  1.71 &  0.17 &  0.29 &  0.01  &  8.4   \\
202484 &  2.32 & 1839  &  1  &  18.3 & 0.012 &  5.79 &  1.63 &  1.03 &  0.13 &  0.37 &  0.01  &  8.4   \\
204595 &  2.87 & 3306  &  1  &  16.0 & 0.027 &  3.48 &  0.04 &  7.58 &  0.08 &  0.04 &  0.001  &  8.0   \\
207524\tablenotemark{g} &  1.99 & 1209 &  1  & 0.318 & 0.020 &  4.60 &  1.04 &  4.21 &  2.22 &  0.08 &  0.04  &  8.2   \\
207888\tablenotemark{g} &  3.33 & 1270 &  1 & 0.215 & 0.013 &  2.72 &  0.42 &  4.89 &  0.75 &  0.06 &  0.04  &  8.1    \\
207982 &  2.00 & 1226  &  1  & 1.58 & 0.010 &  3.82 &  0.29 &  1.53 &  0.13 &  0.16 &  0.01  &  8.3    \\
208000 &  2.58 & 1229  &  1,2,3  & 1.569 & 0.062 & 10.14 &  9.02 &  2.71 & 16.54 &  1.19 &  0.07  &  8.5\\
209962 &  2.23 & 1325  &  1,2,3,4  & 16.6 & 0.089 &  5.69 &  0.48 &  2.24 &  0.07 &  1.40 &  0.02  &  8.5   \\
210730 &  2.23 & 1351  &  1  &  14.0 & 0.025 &  3.00 &  0.08 &  4.07 &  0.11 &  0.09 &  0.004  &  8.2   \\
210885\tablenotemark{g} &  3.08 & 1357 &  1 & 5.34 & 0.031 &  3.87 &  0.06 &  5.71 &  0.08 &  0.08 &  0.003  &  8.1  \\
\enddata
\tablenotetext{a}{MIRI Source ID in the SMILES AGN catalog v0.4.2 published in \citet{Lyu2024};}
\tablenotetext{b}{AGN classification method in \citet{Lyu2024}: 1 = from SED fitting; 2 = X-ray luminosity; 3 = ratio of X-ray to radio; 4 = optical spectrum. Readers should check \citet{Lyu2022b, Lyu2024} for details.}
\tablenotetext{c}{Including slit loss correction;}
\tablenotetext{d}{$\lambda\lambda$ 4969, 5007}
\tablenotetext{e}{$\lambda\lambda$ 6550, 6585}
\tablenotetext{f}{Estimated from O3N2, errors 
$\sim$ 0.2 dex}
\tablenotetext{g}{Dwarf galaxy, $log(M) < 9.5$.}
\end{deluxetable*}

\begin{figure*}  
\centering
 \label{fig:bpt}
\includegraphics[width=0.96\textwidth]{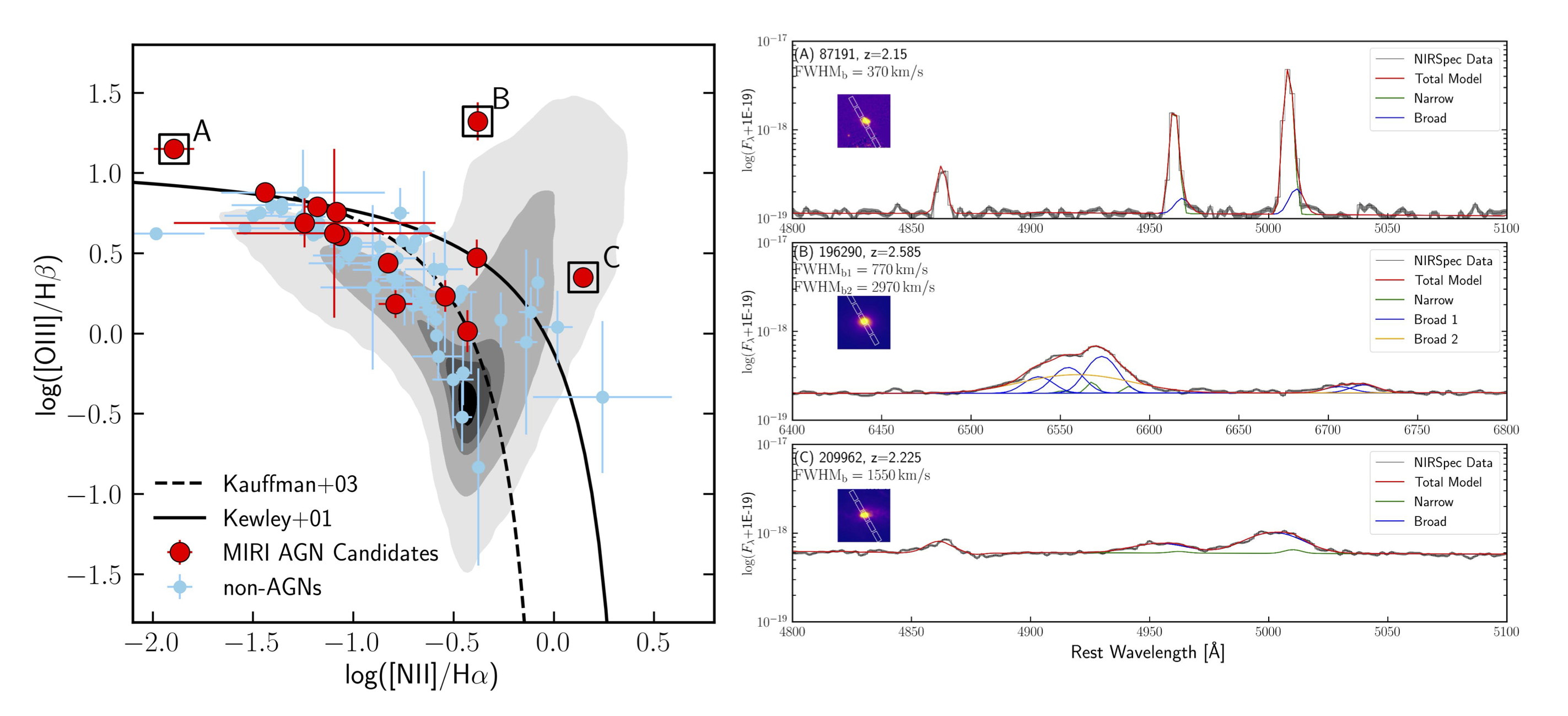}
    \caption{Left: Placement of our sample of galaxies on the BPT diagram with lines of demarcation between star forming and active galaxies from \citet{Kewley2001, Kauffmann2003}. The non-AGN spectra are obtained from other galaxies targeted in the SMILES spectroscopy. Right: Spectra of galaxies ``A'' (87191), ``B'' (209962), and ``C'' (196290) marked on the BPT diagram. ``A'' and ``C'' have lines broadened by outflows, as shown by the similar widths of the [OIII] $\lambda 4959, 5007$ lines to those of the hydrogen lines. ``B'' appears to be a combination of outflow and broad line AGN.}
\end{figure*}

\section{Analysis}

\subsection{BPT Diagram}

Figure~\ref{fig:bpt} shows the galaxies with adequate signal to noise  (SNR$>$3 for  the BPT line ratios) placed on a BPT diagram . The blue points are for non-AGN galaxies with SMILES spectra at a similar redshift range, and they fall in the expected region. Only the 14 AGN galaxies with adequate signal to noise are plotted;  11/14 of these galaxies  have line ratios consistent with star formation being the dominant excitation mechanism\footnote{The diagram shifts modestly at the redshifts of these galaxies, but their values of $z$ range by nearly a factor of three so this is difficult to show simply. However,  all of them remain in the star forming zone since higher redshifts tend to move the lines of demarkation upward and to the right, i.e., away from the points and hence retains their classification as star-forming.}. This is as expected for hosts of highly obscured AGN, where the UV radiation from the AGN is unable to escape  an embedding dense cocoon and excite the interstellar gas outside. 

The exceptions are IDs 196290 (B), possibly 87191 ([A]), and 209962 ([C]). The effect of lower metallicity at high redshift will shift the BPT demarcation line upward putting ID 87191 [A] on the line rather than above it. The classification of this galaxy as star-forming is consistent with its spectrum (Figure~\ref{fig:bpt}), which indicates an inflow or outflow that broadens the [OIII] lines. ID 196290 (B) shows a broad component in its H$\alpha$ line, although the SED implies that much of the output of its AGN is absorbed and re-emitted in the infrared. ID 209962 shows a lightly obscured Type 1 AGN continuum, but the presence of a high velocity [OIII] outflow and the weakness of its H$\beta$ line preclude a sensitive search for a broad line region. 
 To quantify the broad components that we see in the spectra of ID 87191, 196290, and 209962, we re-fit their broad lines ([OIII] for ID 87191 and 209962, H$\alpha$ for ID 196290) using multiple Gaussian components. The narrow component is fixed at the systemic redshift and its width is only restricted within $500\,\mathrm{km/s}$. The broad component is allowed to be blue/red-shifted and must have a width wider than the narrow component. The best fits and the width of the broad component(s) of the three galaxies are shown in the right panel of Figure~\ref{fig:bpt}.

    
\subsection{Star Formation rate estimates}

Our  estimation of star formation rates from H$\alpha$ follows the procedure outlined in \citet{Kennicutt2012}, determining extinction to H$\alpha$ as in \citet{Koyama2015}, Equation 3. This latter approach originated in \citet{Calzetti2001}; updating the extinction coefficients according to \citet{Momcheva2013} gives close agreement to the equation used in \citet{Koyama2015}. 
Details are provided in the Appendix. In some cases, the spectra indicate large Balmer decrements; if these are interpreted in terms of extinction, the derived 
SFR substantially exceeds a plausible level relative to the galaxy main sequence. It may be, for example, that there is H$\beta$ absorption in the stellar continuum, reducing the apparent emission line strength. 
These cases are noted in the footnotes. The SFRs from H$\alpha$ are mildly dependent on metallicity/redshift \citep{Backhaus2024} but not sufficiently to affect our analysis. 

The SFRs  from the spectra  are reported in Table~\ref{sfrlist}, which also lists the SFRs
estimated from fitting the spectral energy distribution using
\texttt{Prospector} \citep{Johnson2021} as carried out  by \citet{Lyu2024} to
identify AGNs in these systems.  The details of the customized
\texttt{Prospector} fitting are described in \citet{Lyu2022b}, Sections 3.1.1
and 3.1.2. and \citet{Lyu2024}, Sections 3.1 and 3.2. We have carried out new
fits to take advantage of improved redshifts and reductions of the photometry.

\begin{deluxetable*}{lccccccc}
\tabletypesize{\footnotesize}
\label{sfrlist}
\tablecaption{SFR estimates and galaxy masses} 
\tablewidth{0pt}
\setlength{\tabcolsep}{1.5pt}
\tablehead{
\colhead {ID} & 
\colhead {z} &
\colhead {SFR (M$\odot$ yr$^{-1}$)}  &
\colhead{err} &
\colhead {SFR (M$\odot$ yr$^{-1}$)}  &
\colhead{err}  &
\colhead {SFR (M$\odot$ yr$^{-1}$)} & 
\colhead {log(M*)}
 \\
\colhead {JADES} &
\colhead {}  &
\colhead {uncorrected}  &
\colhead { } &
\colhead {ext corr} &
\colhead {} &
\colhead {SED fit} &
\colhead{M$_\odot$}
}
\startdata
 87191 &  2.15 &  11.3 &  0.13 &  121.4\tablenotemark{a} &  28.2 &   2.5 & 8.7 \\
113325 &  2.49 &  2.75 &  0.04 &   3.2 &   0.8 &   0.6 & 8.1 \\
194121 &  1.98 &  7.56 &  0.11 &   14.4 &   3.3 &  2.9 & 9.9 \\
194473 &  1.61 &  0.81 &  0.13 &  77 &  30 &  79.9 & 10.7 \\
{\it 196184} &  2.65 &  9.87 &  0.17 & 181\tablenotemark{b} &  43 &  3.0 & 10.1 \\
196290 &  2.59 & 55.7 &  0.8 & 3500\tablenotemark{c} & 900 & 339.9 & 11.1 \\
200525 &  1.24 &  2.68 &  0.04 &   3.9 &   0.9 &  7.3 & 9.3 \\
{\it 201027} &  2.87 &  5.56 &  0.15 &   --- &   --- & 206.8 & 10.8 \\
{\it 202378} &  2.62 &  21.65 &  0.33 &  303\tablenotemark{d} &  72 &  35.8 &9.8\\
{\it 202484} &  2.63 &  55.38 &  1.04 &  348\tablenotemark{e} &   83 &  5.8 & 11.1 \\
{\it 204595} &  2.87 & 59.74 &  0.65 &  99 &   23 &   40.5 & 9.8 \\
207524 &  1.99 &  0.49 &  0.03 &   1.7 &   0.5 &   3.0 & 8.8 \\
{\it 207888} &  3.33 &  1.15 &  0.07 &   1.0\tablenotemark{f} &   0.3 &   2.7 & 8.6\\
207982 &  2.00 &  2.44 &  0.06 &   5.3 &   1.1 &   12.0 & 9.6\\
 208000 &  2.58 &  4.54 &  0.19 & 123.9 &  32.5 &  200.8 & 10.7 \\
209962 &  2.23 & 15.58 &  0.5 &  201 &  48 &  225.4 & 10.8\\
210730 &  2.23 &  28.3 &  0.4 &   32 &   7 &  19.0 & 9.6 \\
{\it 210885} &  3.08 & 23.69 &  0.27 &  52\tablenotemark{g} &  12 &   1.2 & 9.4\\
\enddata
\tablenotetext{a}{The H$\alpha$ line uncorrected for extinction puts this galaxy substantially above the main sequence, indicating it is undergoing a starburst. The huge (and highly uncertain) extinction correction is very likely to be overestimated, since it would put the galaxy a factor of $\sim$ 100 above the MS.}
\tablenotetext{b}{ The extinction corrected H$\alpha$-based SFR puts this galaxy modestly above the main sequence, consistent with a starburst. }
\tablenotetext{c}{Both the uncorrected H$\alpha$-based and the SED-based SFR put this galaxy in the main sequence; the large (and very uncertain) extinction correction to H$\alpha$ would put it an order of magnitude above the MS, which is unlikely. }
\tablenotetext{d}{Both the uncorrected H$\alpha$-based and the SED-based SFR put this galaxy in the main sequence. The large (and very uncertain) extinction correction to H$\alpha$ would put it an order of magnitude above the MS, which is not likely.}
\tablenotetext{e}{The corrected H$\alpha$ SFR (although highly uncertain) puts this galaxy on the main sequence. The value from SED fitting is well below the MS, which is of course possibly correct also.}
\tablenotetext{f}{The extinction is very small, nominally slightly negative.}
\tablenotetext{g}{The uncorrected H$\alpha$ would put this galaxy in the main sequnce and the corrected value would make it a mild starburst. }
\end{deluxetable*}

For z $\gtrsim$ 2.6, even the 6.2 $\mu$m PAH band is at the long end (or beyond) the cutoff of the MIRI 21 $\mu$m filter. Unless high signal to noise has been achieved at 25.5 $\mu$m, the AGN identification can no longer be based on the absence of PAH emission. We have put the IDs of the seven galaxies with z $\ge$ 2.6 in italics in 
Table~\ref{sfrlist}. For these cases, the identification of an AGN is based on the elevated SED at wavelengths $<$ 6 $\mu$m above the expectations for stellar photospheric output and is independent of the SFR estimates. 

\subsection{Individual galaxies}

Here we discuss the galaxies in this study individually. Figure~\ref{fig:brdseds} shows the model fits for all 17 galaxies and is a useful reference for the discussion. Galaxy masses have been determined from the \texttt{Prospector} fits to the photometry. From Figure~\ref{fig:brdseds}, in all but one case the short wavelengths are strongly dominated by the direct stellar emission, which is characterized by many photometric measurements  allowing for accurate synthesis fits. For high mass cases, $log(M) \ge 9.5$, we assume that any infrared excess powered by young stars will have strong PAH emission features, with relatively little emission at wavelengths  short of 6 $\mu$m. We do not make that assumption for dwarf galaxies, which are discussed in detail in Section~\ref{dwarfs}. We use the spectroscopy to check the star formation rates to compare with the ones deduced in the overall SED fitting to test the robustness of any conclusions from the SED fitting for the presence of AGNs. 

{\it 87191}: This is a dwarf galaxy, with $log(M) = 8.7$.  The SFR from the extinction-corrected spectrum  significantly  exceeds that from the SED fit and also falls far above the main sequence (MS); the extinction correction is evidently too large. The SED-determined value puts it in the main sequence. It is object 
[A] in Figure~\ref{fig:bpt}, which nominally puts it barely into the AGN zone; however, if the line ratios are corrected for the typical conditions in the interstellar medium at z = 2.15 by $\sim$ 0.2 dex \citep[][see also \citet{Ubler2023}]{Backhaus2024}, it would come down $\sim$ 0.2 dex in log([OIII]/H$\beta$) and fall right on the boundary. It has a very strong excess above the photospheric emission, by about a factor of two at 2.4 $\mu$m (rest), two orders of magnitude  at 4 $\mu$m (rest), and leveling out toward longer wavelengths. The SED between 3 and 7 $\mu$m is much flatter than that of Haro 11, our metric for hot dust powered by star formation (see Section 4). 
  Its infrared appears to be  
AGN-dominated.

{\it 113325}: This is another dwarf galaxy, with $log(M) = 8.1$. Both the SED fit and the spectrum indicate it has a very low level of star formation {\bf ($<$ 4 M$_\odot$ yr$^{-1}$)}, within or above the MS. Its excess is both too large and too hot for a star forming galaxy (see Section~\ref{sec:JHK}); Figure~\ref{fig:brdseds} shows that its excess is significant at wavelengths short of 2 $\mu$m (conveniently marked by the Pa $\alpha$ line at 1.9 $\mu$m).  Its infrared is 
AGN-dominated. 

{\it 194121}:
The SFR {\bf ($\sim$ 14 M$_\odot$ yr$^{-1}$)} puts this galaxy in the MS. The infrared for this galaxy is PAH-dominated, although there is a small excess above the purely PAH fit between 4 and 6 $\mu$m (rest) that suggests a small contribution from an AGN.  

{\it 194473}: This galaxy has a fairly high level of star formation as indicated by both the SED fit and the spectrum  ($\sim$ 80 M$_\odot$ yr$^{-1}$), putting it in the MS. It is also relatively bright and massive, with $log(M) = 10.7$. These parameters indicate it should have strong PAH emission, but the data show the PAH contribution to the infrared to be modest, indicating that it is AGN-dominated. 

{\it 196184}: This galaxy is at z = 2.65, but we have a measurement at 25.5 $\mu$m. At its mass $log(M) = 10.1$, it should be PAH dominated if it were predominantly star forming and thus have low excess at 5 $\mu$m and shorter wavelengths.  Although it has a high level of star formation ($\sim$ 180 M$_\odot$ yr$^{-1}$), it has a very substantial excess between 3 and 6 
$\mu$m, and only a small hint of a PAH feature, requiring  the presence of an AGN. 

{\it 196290}: The SFR is very high according to both the spectrum and SED fit { ($\sim$ 100 M$_\odot$ yr$^{-1}$), and the galaxy is very massive, with $log(M) =  11.1$. Nonetheless its output is  overwhelmed by a hot excess from 2 - 6 $\mu$m (rest) with only a weak suggestion of PAH features, indicating an obscured AGN.  Its spectrum (Figure~\ref{fig:bpt}) shows very strong outflows and a possible  weak broad line from an AGN. It falls in the AGN region in the BPT diagram, indicating that the AGN is only partially embedded.

{\it 201027}: The H$\beta$ line is not detected sufficiently well for an extinction correction, but the SED fit indicates a very high level of star formation {\bf ($\sim 200$ M$_\odot$ yr$^{-1}$).} The galaxy is massive, with $log(M) = 10.8$ and should be PAH dominated, with little emission from stellar-heated dust short of 5 $\mu$m. However, there is a large excess between 3 and 6 $\mu$m, indicative of an AGN.

{\it 202378}: The level of star formation is moderately  high from the spectrum  ( $\sim 35$ M$_\odot$ yr$^{-1}$); we have discounted the extinction-corrected H$\alpha$-based SFR because it is implausibly high relative to the MS. We have a 25.5 $\mu$m detection, that should pick up PAH emission , but the SED fit (Figure~\ref{fig:brdseds}) shows a significant hot excess above the expected PAH contribution, indicating an AGN. 

{\it 202484}: The extinction-corrected H$\alpha$ would put this galaxy toward the top of the MS.  Both the spectrum and SED
 fit indicate a moderate level of star formation with no need for an AGN. 
In \cite{Lyu2024}, this source was identified as an obscured AGN assuming a redshift of $z=2.32$ from the 3D-HST grism spectrum (grade A). However, our MSA observation yielded a confident spec-$z$ value at $z=2.63$. With the increased redshift, the new SED analysis does not confidently suggest the existence of an AGN. Nevertheless, some excess emission above the galaxy template at $\lambda$4 --6 $\mu$m can be still seen, indicating the possible existence of a weak AGN.

 {\it 204595}: The estimate of the SFR from the SED fit agrees reasonably well with that from the spectrum   ( $\sim 100$ M$_\odot$ yr$^{-1}$), putting it near the top of the MS. This galaxy is at a redshift where the PAH bands are shifted out of the MIRI range, but with $log(M) = 9.8$ it should be PAH dominated with a resulting weak excess in the 4 - 6 $\mu$m range. Nonetheless,  There is a  strong mid-infrared AGN component.

 {\it 207524}: This is a dwarf galaxy with $log(M) = 8.81$. With the update to the redshift and improved photometry \citep{Alberts2024}, what appeared to be a modest mid-infrared excess in \citet{Lyu2024} has disappeared. 

{\it 207888}: This is a dwarf galaxy with $log(M) = 8.65$. Its star formation rate is low {\bf (only $\sim 1.5$ M$_\odot$ yr$^{-1}$),} from both the spectrum and SED fit, but consistent with the MS. It has a very strong mid-infrared component starting from 1.5
$\mu$m that must be associated with a buried AGN  (see Section~\ref{sec:JHK}).

{\it 207982}: The level of star formation is modest  ($\sim 7$ M$_\odot$ yr$^{-1}$), showing the galaxy lies in the MS. There is a substantial excess in the 5 $\mu$m range, confirming that the mid-infrared excess arises from a buried AGN.

{\it 208000}: Both the spectrum and the SED fit indicate ULIRG-level star formation {\bf ($\sim 130$ M$_\odot$ yr$^{-1}$).} The longest wavelength MIRI bands  (including 25.5 $\mu$m) are dominated by the PAH emission, but at shorter wavelengths the photometry requires an AGN component to boost the output above the pure PAH spectrum.

{\it 209962}: This galaxy  has ULIRG-level star formation from both the spectrum and the SED fit ($\sim 200$ M$_\odot$ yr$^{-1}$). It also has a lightly obscured AGN whose output is dominant at wavelengths $>$ 1.5 $\mu$m. It has a very high-velocity outflow \citep{Zhu2025}  Figure~\ref{fig:bpt}) and the H$\beta$ detection is weak; these characteristics may hide evidence for a broad line region in our spectrum.

{\it 210730 }: Both the spectrum and the SED fit indicate a moderate level of star formation  ($\sim 30$ M$_\odot$ yr$^{-1}$), which puts it in or near the MS. The longer wavelength MIRI bands are dominated by PAH emission but an additional AGN component is required to fill in at the shorter wavelengths.

{\it 210885}: This is a dwarf galaxy with $log(M) = 9.4$. The spectrum indicates a higher rate of star formation {\bf ($\sim 50$ M$_\odot$ yr$^{-1}$)} than the estimate from SED fitting and the excess attributed to an AGN is seen only at $\lambda > 3 \mu$m, where it could be powered by hot stars, so the identification of an AGN is ambiguous. 

In summary, we selected a sample with a variety of characteristics to test use of SED fitting  to identify obscured AGN in massive galaxies ($log(M) > 9.5$ ). The results generally confirm the approach used by \citet{Lyu2024}, except where there are errors in redshift. The galaxies ID 87291,  196184, 202484, and 207524 had such errors. For 207524, new modeling at z = 1.99 from our spectrum (instead of a claimed spectroscopic redshift of 1.13) was responsible for the ``disappearance'' of the obscured AGN compared with \citet{Lyu2024}. For 87291, the original modeling was for z = 2.48 from Zeasy (i.e., photometric), but our spectroscopy gave z = 2.15. Since the correct value is lower, the implications for AGN identification are minor. For 196184 and 202484, the corrected redshifts are increased, 2.08 to 2.65 and 2.32 to 2.63, respectively, but the changes did not undermine the identifications of AGNs. For both of these latter galaxies, the original redshifts were spectroscopic, but our full high signal-to-noise sampling with NIRSpec leaves no doubt in the new values. 

Other than the issues with redshift, the results of this study generally confirm those from SED fitting in \citet{Lyu2024}.


\begin{figure*}
    
   \centering
\includegraphics[width=0.45\textwidth]{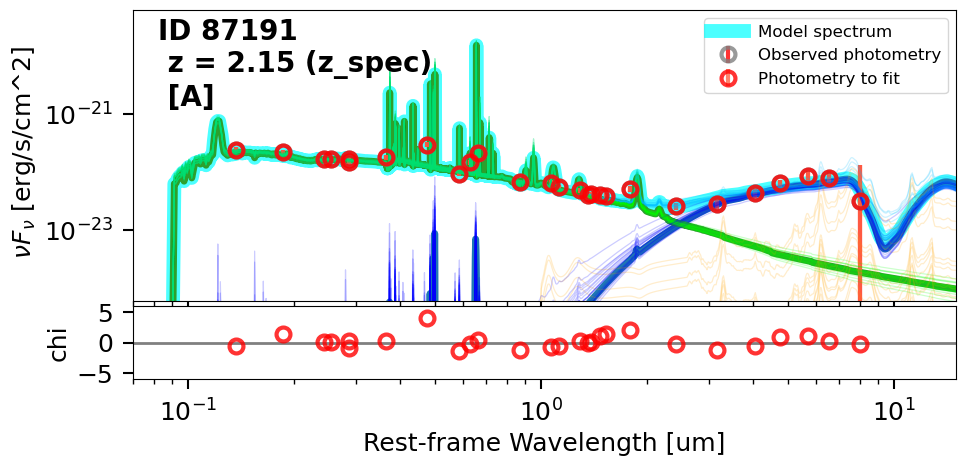}
\includegraphics[width=0.45\textwidth]{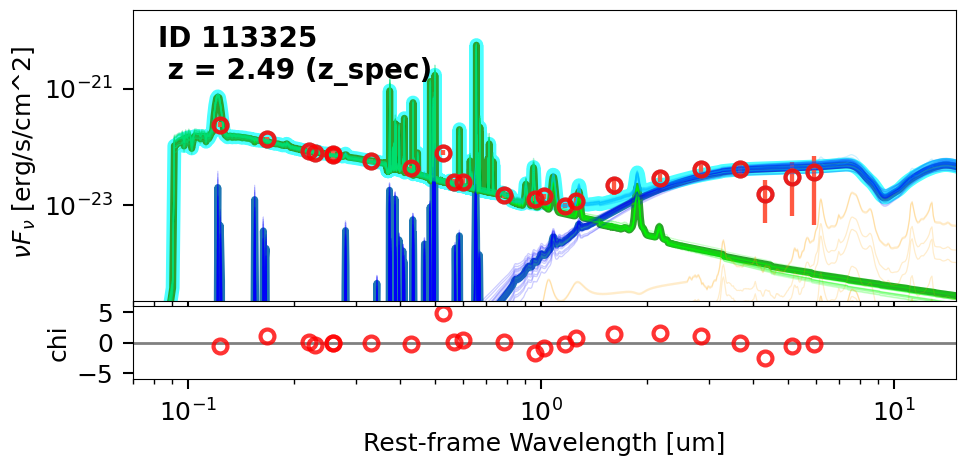}
\includegraphics[width=0.45\textwidth]{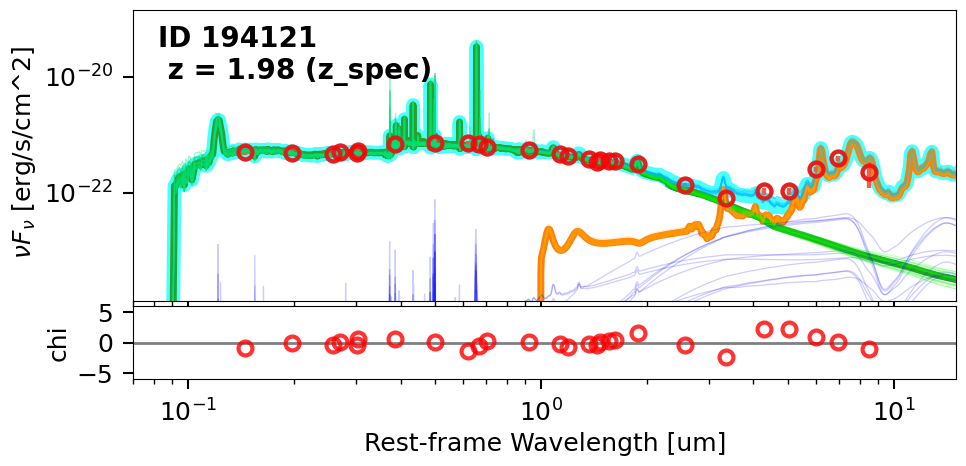}
\includegraphics[width=0.45\textwidth]{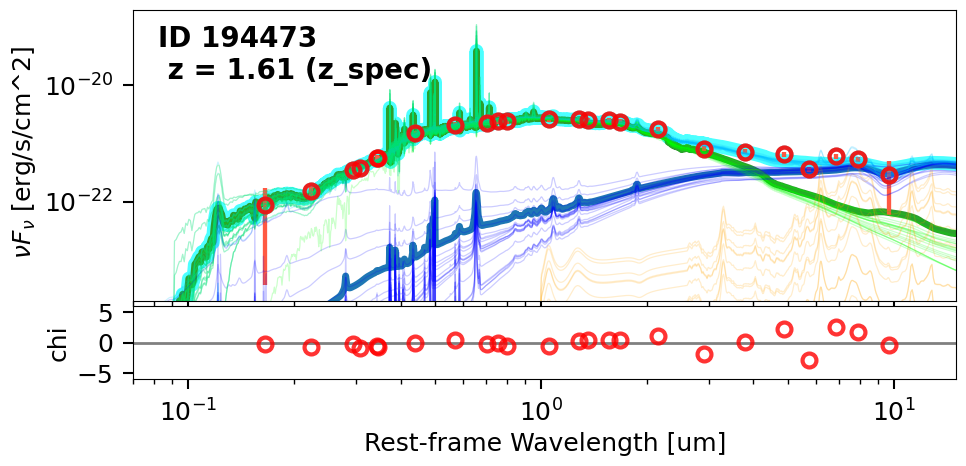}
\includegraphics[width=0.45\textwidth]{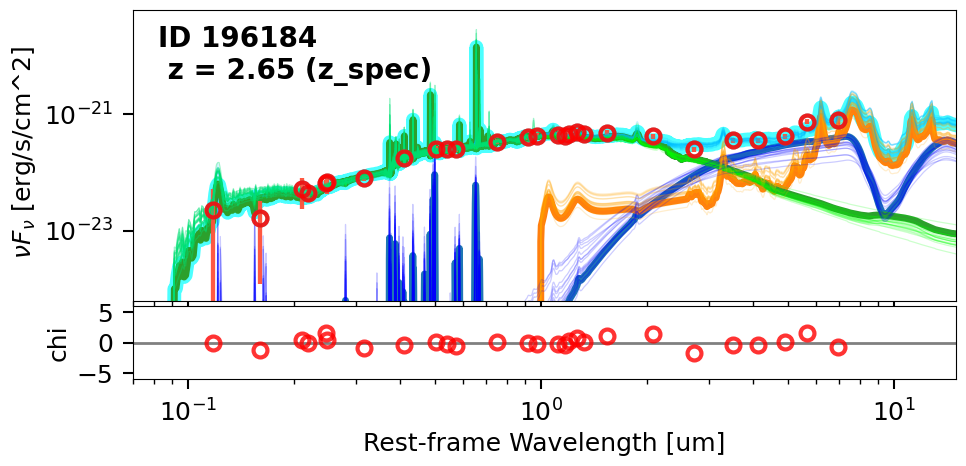}
\includegraphics[width=0.45\textwidth]{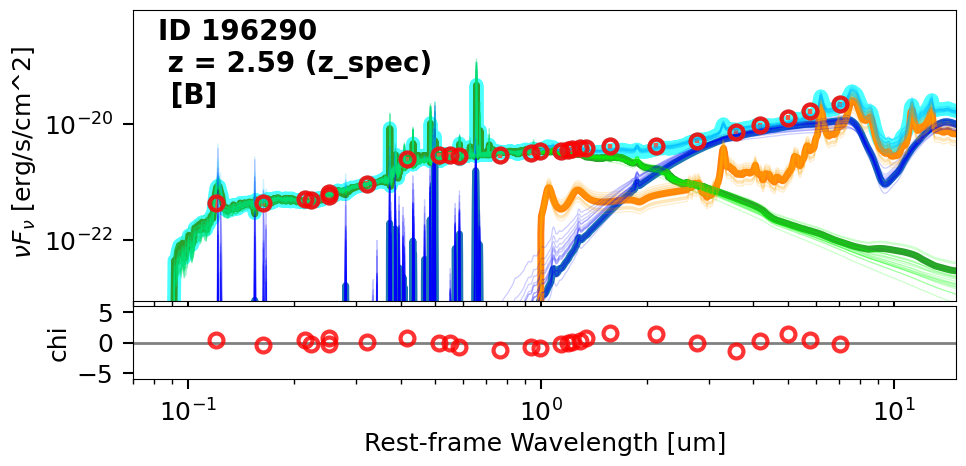}
\includegraphics[width=0.45\textwidth]{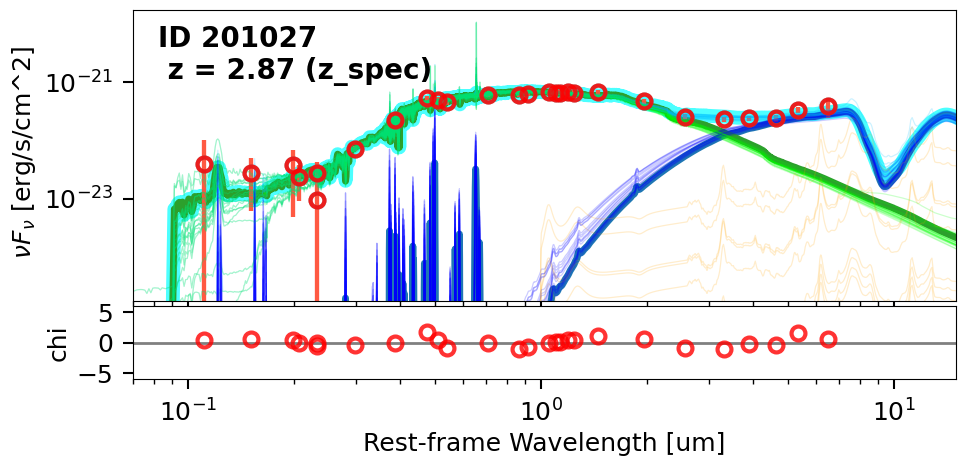}
\includegraphics[width=0.45\textwidth]{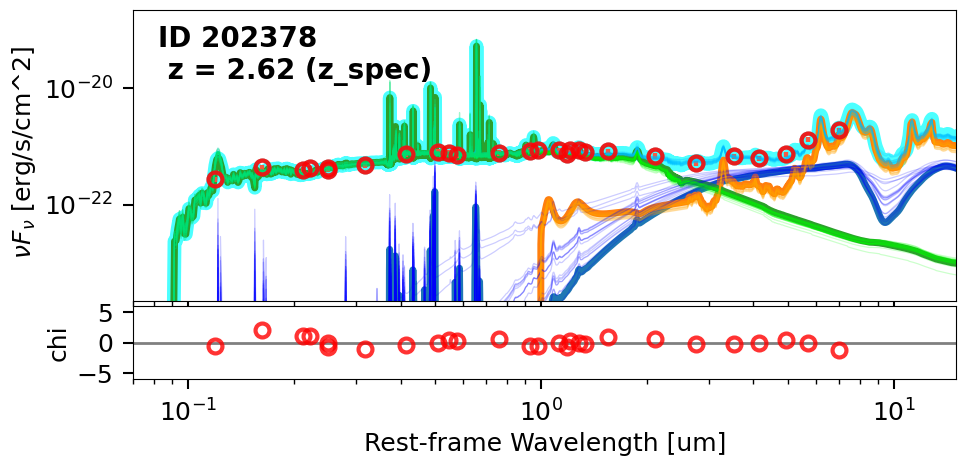}
\includegraphics[width=0.45\textwidth]{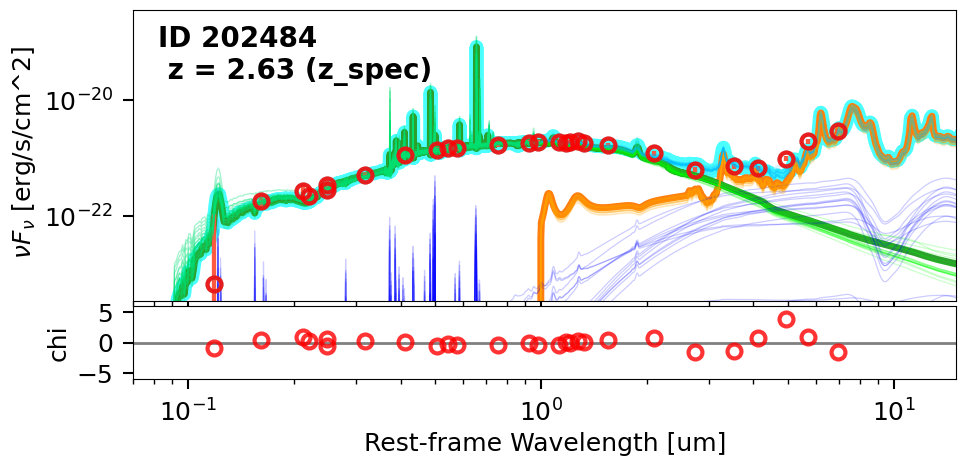}
    \caption{ Decomposition of photometry into stellar (green lines), AGN (purple lines), and stellar-powered infrared excess (orange lines). We present the galaxy ID and redshift in the top-left corner of each panel. To demonstrate the model degeneracy, we use the light lines to indicate a range of fits that would be acceptable to emphasize that they have little effect on the decision of whether an embedded AGN is present.} 
    \label{fig:brdseds}
\end{figure*}

\begin{figure*}
    \figurenum{2}
   \centering

\includegraphics[width=0.45\textwidth]{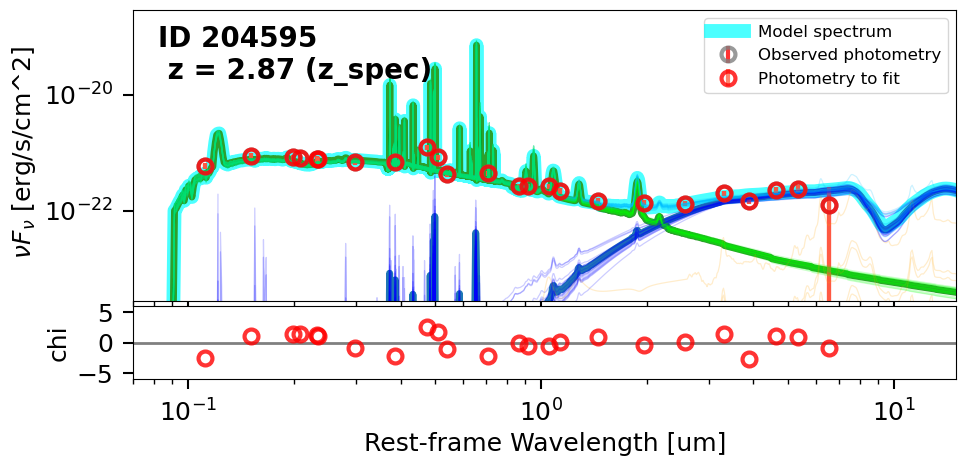}
\includegraphics[width=0.45\textwidth]{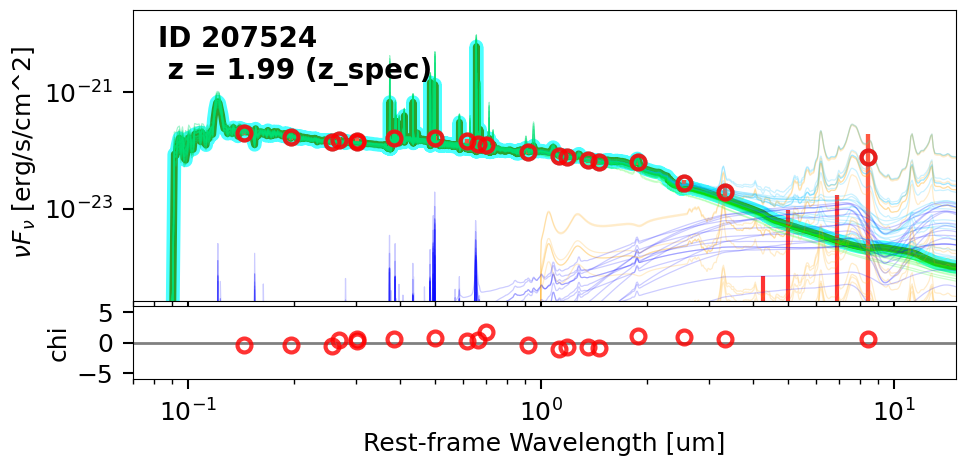}
\includegraphics[width=0.45\textwidth]{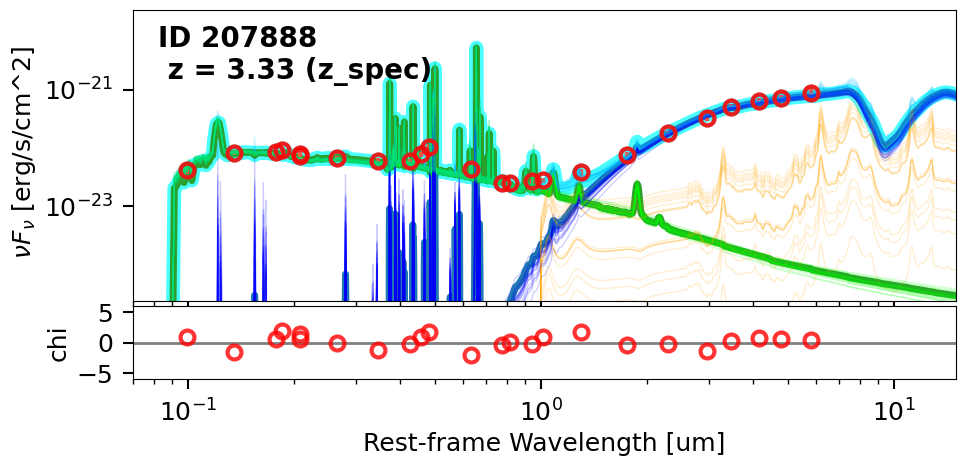}
\includegraphics[width=0.45\textwidth]{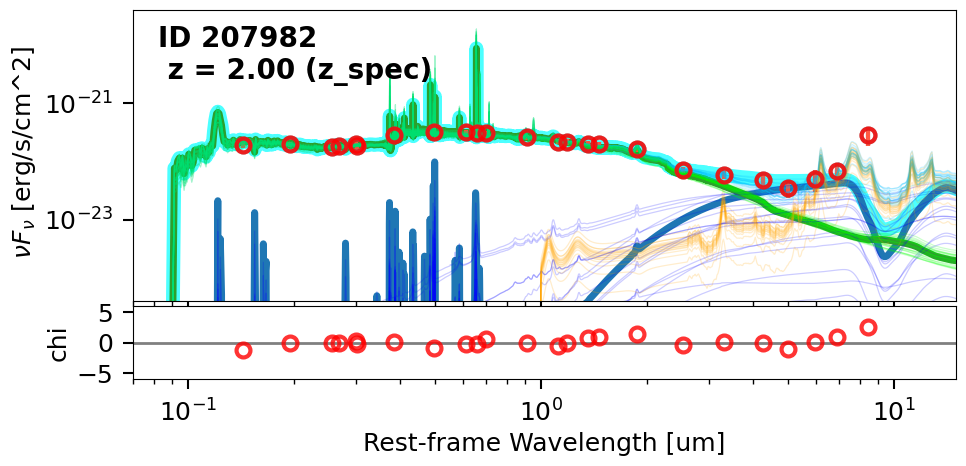}
\includegraphics[width=0.45\textwidth]{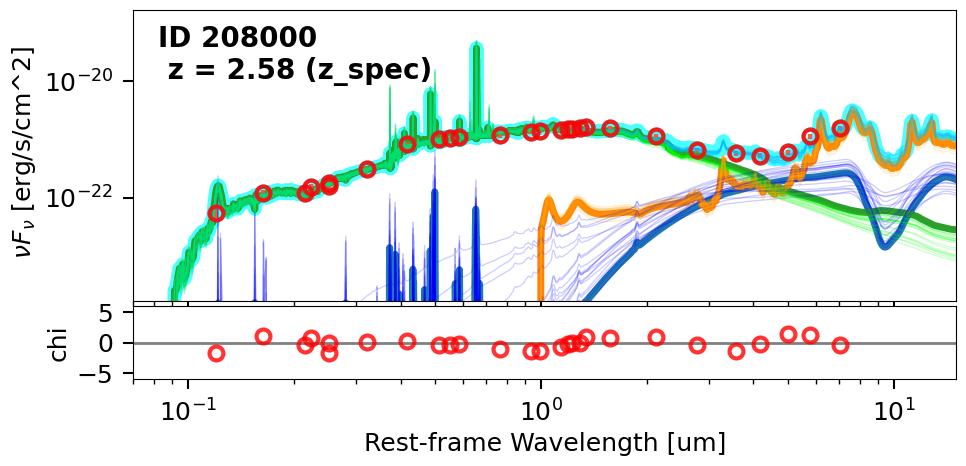}
\includegraphics[width=0.45\textwidth]{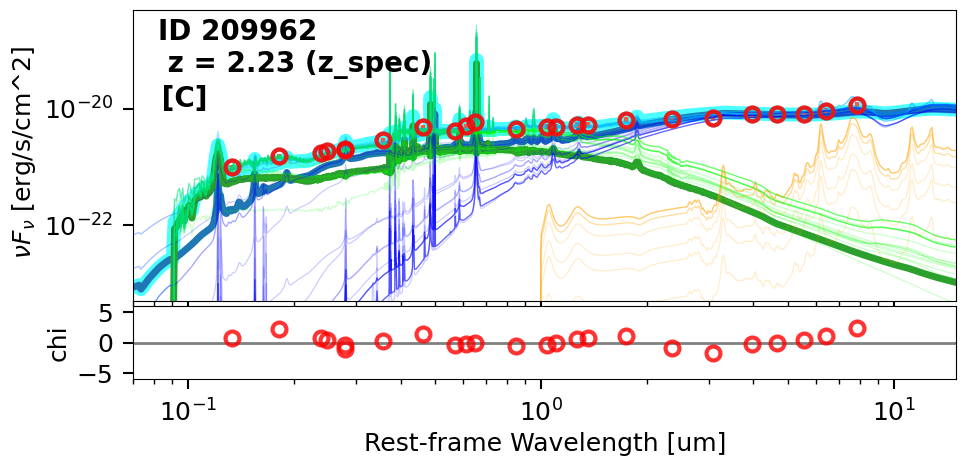}
\includegraphics[width=0.45\textwidth]{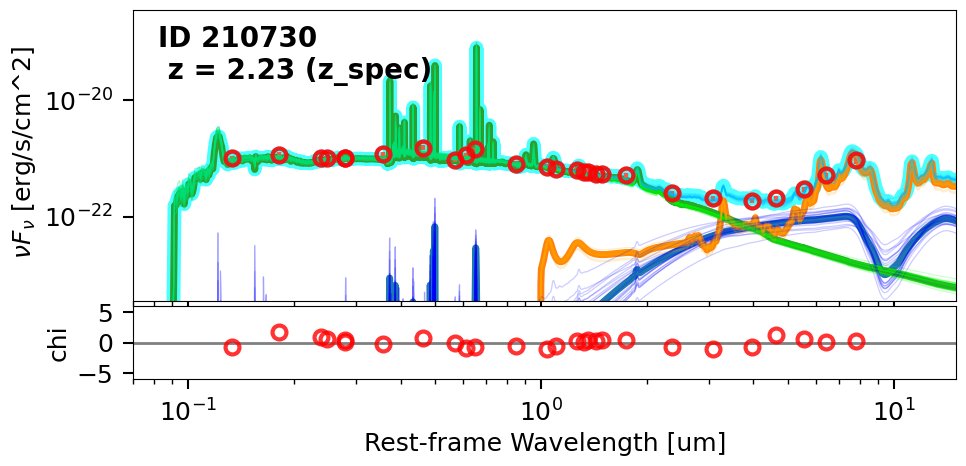}
\includegraphics[width=0.45\textwidth]{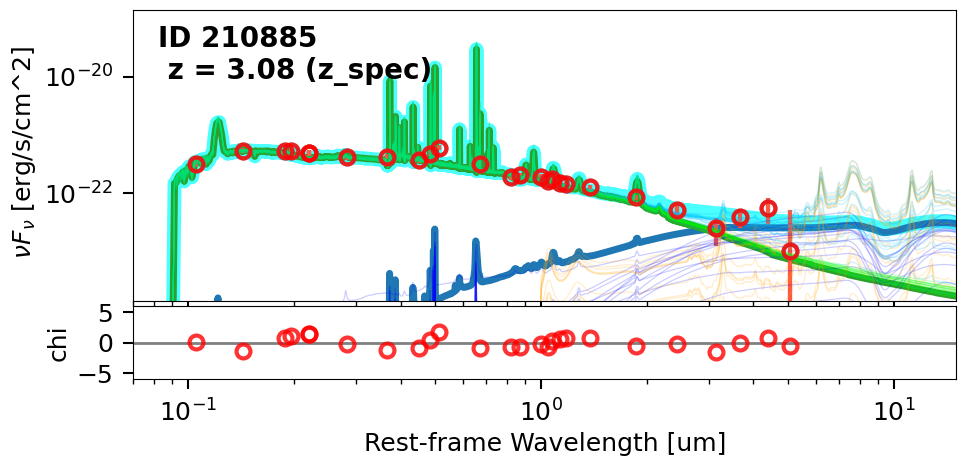}
    \caption{(Continued.)}
    \label{fig:brdseds2}
\end{figure*}

\section{AGN in dwarf galaxies with warm dust}
\label{dwarfs}

We now discuss the selections of AGNs in dwarf galaxies using just SED
information, moving beyond the cases with spectra. As shown in Section 3.1, even
for massive galaxies, only a small fraction of heavily obscured AGNs can be
revealed by optical spectral properties \citep[see also][]{Lyu2022b}. Thus, we
need to rely on photometric data to provide a more complete AGN sample for the
dwarf galaxy population. The major challenge is that, in contrast to massive
galaxies whose star-forming dust emission spectra are well characterized, strong
PAH features are no longer universal at the low metallicities of most dwarf
galaxies, requiring a re-evaluation of the selection strategy.

Specifically, for low-metallicity systems, dust emission powered purely by
hot stars can reach higher temperatures than at solar metallicity
\citep[e.g.,][]{Draine2007}, because of the hotter stellar UV continua and the
different grain properties in low metallicity environment. As a result, low
metallicity star-forming galaxies can be confused with AGNs in the near- to
mid-IR, as suggested by \citet{Hainline2016, Sturm2025}. However, the number
of potentially AGN-impersonating galaxies (hereafter AIGs) in the sample of
\citet{Hainline2016} is a very small fraction, as they pointed out. They took
their sample from \citet{Reines2013}, who found 136 AGN or composite galaxies
from a sample of 25974 in the NASA-Sloan Atlas
\footnote{\url{https://nsatlas.org/}}. In their study, \citet{Hainline2016} drew
14013 cases with good WISE detections  (median redshift $z = 0.03$), of which 10 fell within the
\citet{Jarrett2011} AGN zone in a W2-W3 vs. W1-W2 plot, i.e., only 0.07\% of the sample has 
AGN-like WISE colors. More liberally, if the \citet{Stern2012} line was adopted, 31 sources matched the suggested AGN WISE color, still only
0.2\% of the whole population. Other infrared searches for AGN in dwarf galaxies return similarly low identification rates \citep[e.g.,][]{Sartori2015, Mezcua2024}. Such values are probably upper limits, as we will show in Section~\ref{sec:BPTAGNs}
by estimating the fraction of the AIGs that, with more complete spectroscopy,
actually {\it do} have AGNs. We will then discuss galaxy templates that are better suited to fit 
the low-metallicity galaxy population during AGN identification.

\subsection{Spectroscopic evidence for AGNs}
\label{sec:BPTAGNs}

In evaluating the overall possibility of mimicking AGNs, it is desirable to
assess how likely a dwarf galaxy with warm dust will {\it not}
contain an AGN. To do so, we have focused on the 11 galaxies with AGN-like color and WISE W1
$\gtrsim$ 14 in \citet{Hainline2016}. The data are less complete and of lower
quality for the fainter galaxies, so this selection can give us a small unbiased
sample with relatively good measurements. All eleven have reasonably high
signal to noise measurements in WISE W3 and W4 and six are detected in the IRAS faint source
catalog at 60 $\mu$m. 

Table~\ref{agnlist} shows the result of a literature search for spectroscopic
indications of AGN in these 11 galaxies. Of the ten where measurements could be
found, six appear to have AGN or are composite. That is, a substantial fraction
are not pretending to have AGNs, they actually {\it have} AGNs. The role of these AGNs
in the galaxy infrared emission is not well determined, but they are likely to add flux
at the shorter wavelengths and increase the apparent temperature of the ``hot''
dust. This also suggests that the fraction of false AGN identifications based on mid-IR colors due to
very hot dust is likely to be below 0.2\%. 

\begin{deluxetable*}{llll}
\tabletypesize{\footnotesize}
\label{agnlist}
\tablecaption{Indications of AGN in Dwarf Galaxies} 
\tablewidth{0pt}
\setlength{\tabcolsep}{1.5pt}
\tablehead{
\colhead {SDSS ID} & 
\colhead {Type} &
\colhead {Indicator}  &
\colhead {Source} 
}
\startdata
 J005904.10+010004 & SF &   BPT, no X-rays
 & \citet{Sturm2025, Latimer2021} \\
J032224.64+401119.8 &  AGN
 & BPT, X-rays & \citet{Reines2013, Latimer2021}   \\
J083538.40-011407.0 &   AGN\tablenotemark{a}  & BPT & see note \\
J090613.75+561015.5 &  AGN & BPT, brd H$\alpha$, X-rays & \citet{Reines2013, Kawasaki2017, Aravindan2024, Latimer2021}  \\
J095418.15+471725.1 &  AGN & BPT, X-rays & \citet{Reines2013, Latimer2021} \\
J123948.06-171753.9 &  29\% AGN\tablenotemark{b} & BPT & \citet{Chen2022}  \\
J133245.62+263449.3 &  AGN & BPT, X-rays & \citet{Reines2013, Latimer2021} \\
J173501.25+570308.8 & composite\tablenotemark{c}  & brd H$\alpha$ & \citet{Mezcua2024}  \\
J225515.30+180835.0 &  -- & -- & no spectrum available  \\
J231544.60+065439.0 &  SF & BPT & \citet{Wang2018}  \\
J233244.60-005847.9 &  SF &  BPT & \citet{Sturm2025}  \\
\enddata
\tablenotetext{a}{\citet{Chen2022} give a probability of 30\% that there is an AGN, but examination of the 6DF spectrum shows that nomionally log([OIII]/H$\beta$) $\sim $ 1.6, with H$\beta$ weakly if at all detected. This would put the galaxy close to point [B] in Figure~\ref{fig:bpt}. The 3$\sigma$ lower limit on log([OIII]/H$\beta$) is $>$ 1.08, while log([NII]/H$\alpha$) $\sim$ -0.2, still putting the galaxy into the AGN realm.  }
\tablenotetext{b}{\citet{Chen2022} give a probability of 29\% that there is an AGN.}
\tablenotetext{c}{\citet{Latimer2021} find the X-ray properties are ambiguous and classify them as ``XRB/ULX
 or AGN,'' where XRBs are high-mass X-ray binaries and ULXs are ultraluminous X-ray sources located off the nucleus. For J173501.25+570308.8, \citet{Latimer2021} have no HST image so the X-ray source could be nuclear, which along with its high luminosity \citep{Chen2022} increases the likelihood that it contains  an AGN. }
\end{deluxetable*}

\subsection{A Broader Perspective}

\subsubsection{Behavior of dwarf galaxies at high redshift}

The sample of \citet{Hainline2016} is all at relatively low redshift, whereas we want to use SED fitting to find AGNs up to the highest redshifts (z $\le$ 3) where extensive photometric MIRI surveys for AGN have been conducted. If the incidence of very high SFR density is much greater at these redshifts, there would be additional false identifications. 

Although the star formation rates in massive galaxies decline substantially from cosmic noon to the present, this effect is greatly reduced for dwarfs with a total drop of only $ \times 3$ from z $\sim$ 3 to 0.1 \citep[][Figure 4]{Forster2020}. In addition, the size evolution over this interval is modest \citep[e.g.,][]{Mowla2019, Euclid2025}. We therefore expect only a modest increase in the fraction of dwarf galaxies with a very high density of star formation. 

\subsubsection{Detailed Dwarf 
Galaxy Properties}

Most galaxies studied in \citet{Hainline2016} are faint and relatively
little is known about them. To probe the behavior over a sample with better  determined properties,  we consider the members of the Dwarf Galaxy
Survey (DGS) and KINGFISH, which have been studied in detail by \citet{Remy2013,
Remy2015}.   So
far as is known, none of these galaxies have an AGN of  sufficient power to influence its general properties
\citep{Remy2015}.

\citet{Remy2015} provide aperture-corrected photometry in both the IRAC and WISE bands. Figure~3 shows these galaxies on a WISE color-color plot, similar to that used
in \citet{Hainline2016}. As they found, few examples appear to be AGN
candidates. We have labeled galaxies of interest; they fall into two categories,
(1) very low metallicity (I Zw 18, HS0017+1055, and SBS0335-052 with 12+log(O/H)
of 7.14 , 7.63, and 7.25 respectively \citep{Remy2013}) or (2) with very compact
nuclear star-forming regions (NGC 5253 \citep{Smith2020}, II Zw 40 \citep{Kepley2014} and Haro 11
\citep{Derossi2018}). However, this diagram provides limited insight for what might be found above z = 1, since the W3 band shifts to longer observed wavelengths than
those of any suitable observations.

Therefore, in Figure~4, we show the IRAC color-color plot proposed by
\citet{Stern2005}, with the longest band at 8 $\mu$m. \citet{Donley2012}
demonstrate the validity of this diagram for AGN selections up to z $\sim$ 3. They also show that
the selection region lies along a line, which we show in the figure. This plot adds
Tol1214-277, another very low metallicity galaxy with 12+log(O/H) = 7.52
\citep{Remy2013}. Of the seven galaxies in the sample with 12+log(O/H) $\le$
7.63, four are within the AGN zone. NGC 930 and II Zw 40 are at the edge of the
AGN zone, but they are far from Donley's optimized zone, so we do not count them
as indicated AGNs. 

Both color-color plots show that galaxies with very low metallicity, i.e. $\le 0.1  Z_\odot$, can have colors, and hence SEDs, that mimic AGNs. They also reinforce the suggestion from \citet{Sturm2025} that very compact nuclear starbursts can have infrared colors similar to those of obscured AGN.

\begin{figure}  
   \centering
 \label{fig:WISEcolor}
\includegraphics[width=0.4\textwidth]{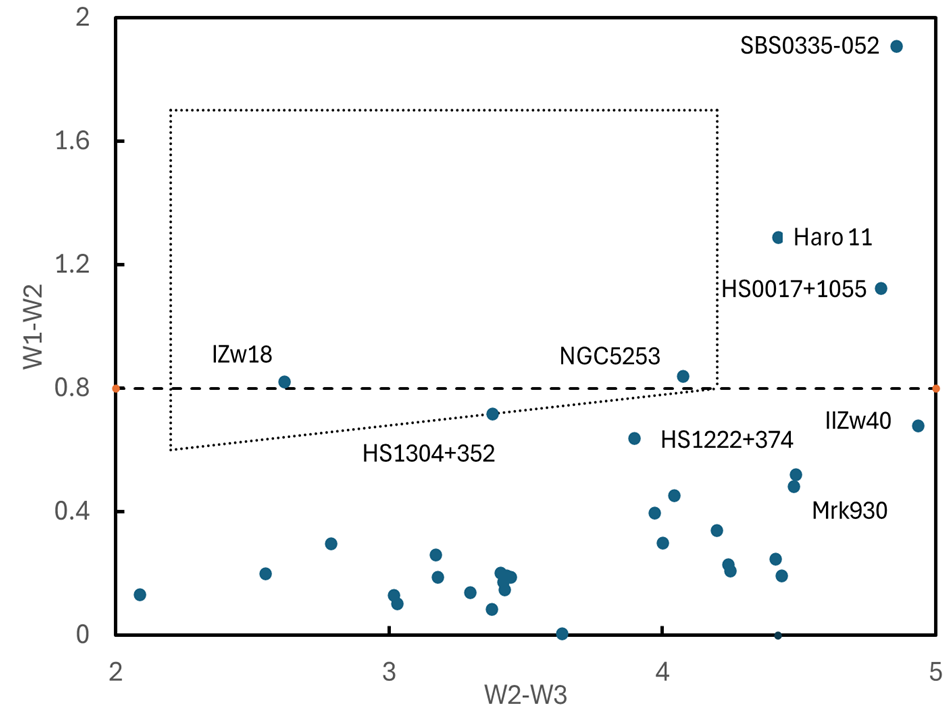}
    \caption{WISE color-color plot similar to those used in \citet{Hainline2016} but for the sample of \citet{Remy2013}. The dotted line outlines the AGN zone from \citet{Jarrett2011} and the dashed line is the W1$-$W2 criterion proposed by \citet{Stern2012}}
\end{figure}

\begin{figure}  
   \centering
 \label{fig:IRACcolor}
\includegraphics[width=0.4\textwidth]{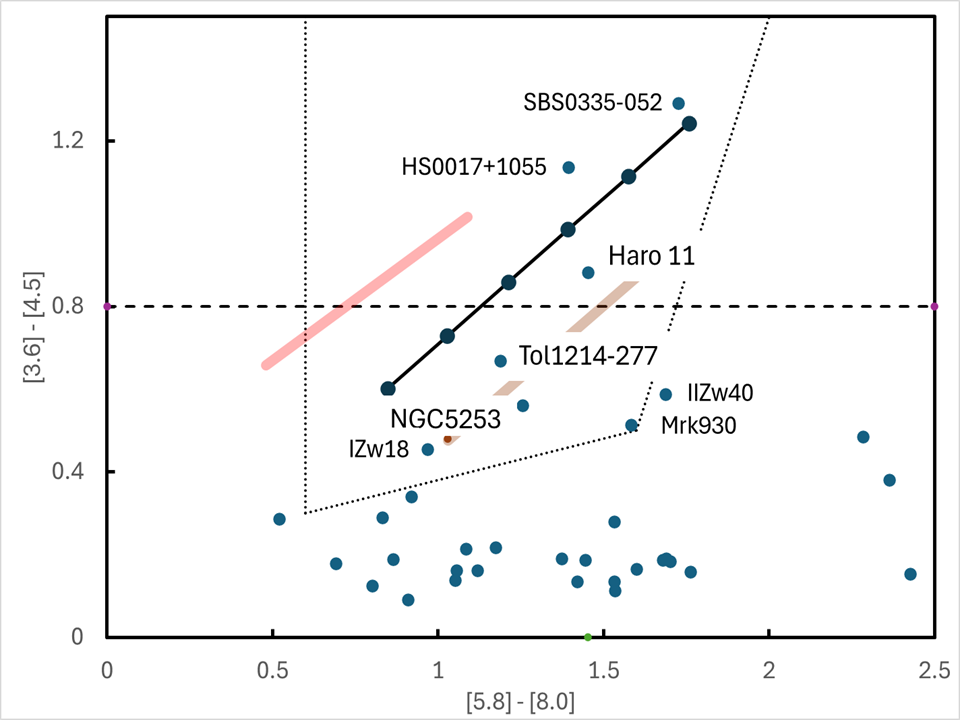}
    \caption{IRAC color-color plot for the sample of \citet{Remy2013}. The dotted line outlines the AGN zone from \citet{Jarrett2011} and the dashed line is the W1-W2 criterion proposed by \citet{Stern2012}. The dotted line is the refined relation from \citet{Donley2012} and the broad red lines represent the maximum scatter around this relation for 5\% errors, which are typical for the measurements plotted. }
    \label{fig:iraccolors}
\end{figure}

\subsubsection{$JHK_S$ colors}  \label{sec:JHK}
We can extend the above investigation to shorter wavelengths using the $JHK_S$ colors of the galaxies. We have used the photometry in \citet{Remy2015} to determine standard colors for low metallicity galaxies. We obtain $J-H = 0.60 \pm 0.03$,  $H-K_S = 0.25 \pm 0.03$, and $K_S-W1 = 0.34 \pm 0.04$.  We compare these colors with those for the sample with W1-W2 excesses in \citet{Hainline2016}.  We have used the standard 2MASS point source photometry to focus on the centers of the galaxies. The colors are shown in Figure~\ref{fig:JHKcolors}.

Three galaxies stand out in the figure. Number 1 is J123948.06-171753, spectroscopically a star forming galaxy with a modest probability of having an AGN (see Table~\ref{agnlist}). Despite its relatively red $J-H$ and $H-K_S$ colors, its $K_S-W1$ is 0.45, hardly different from the standard stellar-dominated color.  If the red JHK colors were associated with either hot dust or an 
AGN, we would expect a much redder $K_S-W1$. The nature of its behavior is not clear. Galaxies 2 \& 3 contain AGNs (Table~\ref{agnlist}) so their modest red offsets are not necessarily associated with emission by star-heated dust. All the other galaxies are consistent with the standard $J-H$ = 0.60  -- and these three cases are also within the errors. This shows that the excess emission identified through the extreme W1-W2 color is gone at 1.6 $\mu$m, the effective wavelength of the $H$ band. That is, the warm dust is not so hot as to emit significantly at this wavelength. 

In comparison, for most of the galaxies there {\it is} excess emission in the $K_S$ band, showing that the warm dust identified through W1-W2 does have detectable emission at its effective wavelength, 2.16 $\mu$m.

\begin{figure}  
   \centering
    
\includegraphics[width=0.4\textwidth]{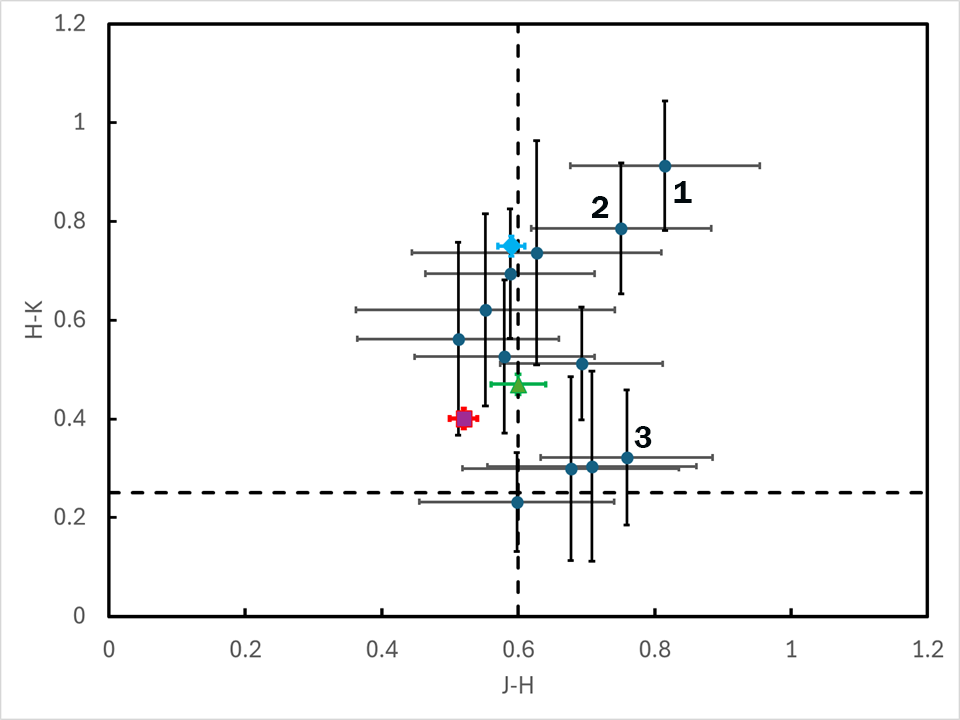}
    \caption{$H-K_S$ vs. $J-H$ for the subset of AIGs and Haro 11, NGC 5253, and II Zw 40. The standard colors for low metallicity galaxies are shown as dashed lines. An AGN-like excess would lie above  the vertical line and to the right of the vertical one. In this case we use the 2MASS point source photometry to focus on the galaxy nuclei and also because it has higher signal to noise than the extended source photometry. The position of Haro 11 is shown as a blue diamond, photometry from \citet{Vader1993}; photometry for NGC 5253 is a red square, from the 2MASS extended objects catalog 14$''$ beam; and II Zw 40 is a green triangle from \citet{Thuan1985}. ``1'' is J123948.06-171753.9, ``2'' is J032224.64+401119.8 and ``3'' is J133245.62+263449.3, cases that are discussed in the text.}
\label{fig:JHKcolors}
\end{figure}

\subsection{SED of Haro 11 as a template}
Previous work has employed the SED of Haro 11 as a template to distinguish embedded AGNs in dwarf galaxies from dust powered by stars \citep[e.g.,][]{Lyu2024}. Most of the  infrared emission of Haro 11 originates from central knot B, which
appears to be a scaled up version of the central clusters in the four AIGs
studied by \citet{Sturm2025}: diameter $\sim$ 140 pc, young stellar mass $\sim 3
\times 10^8$ M$_\odot$ \citep{Derossi2018}. Its very  powerful star formation, $L \sim 5 \times 10^{10}$ L$_\odot$, results in an average  luminosity density in its interstellar medium $\gtrsim 10^4$ L$_\odot$ pc$^{-3}$, one of the highest values known among star forming galaxies. Other examples with similarly high luminosity densities have similar infrared SEDs, e.g., NGC 5253, II Zw 40.

All three examples lie near or in the AGN color range in Figure~\ref{fig:iraccolors}, making them representative AIGs. The $JHK_S$ colors of these three galaxies in Figure~\ref{fig:JHKcolors} show that their dust is sufficiently hot to make $H - K_S$ slightly red compared with standard colors, particularly true for Haro 11. These results show that Haro 11 is representative of the most extreme star forming dwarf galaxies in terms of its very warm interstellar dust and relatively ``hot'' SED. 
Haro 11 is sufficiently luminous that a detailed spectrum is available.  This well sampled and high signal-to-noise spectrum is a useful template for this type of behavior.  It is very likely that any galaxy with a warmer infrared SED hosts an AGN.

  \begin{figure}  
   \centering

\includegraphics[width=\linewidth]{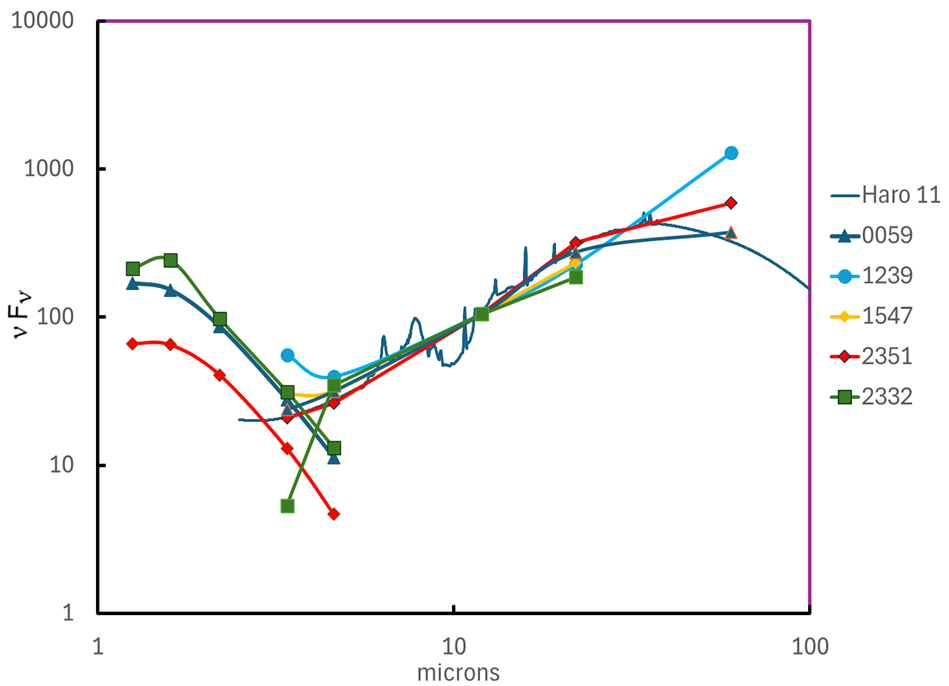}
    \caption{Comparison of the SEDs of non-AGN galaxies with that of Haro 11, all normalized at 
    W3. The range from 4 through 12 $\mu$m for the dwarf galaxies is matched well by the SED of Haro 11. (We do not show J160135.95+311353.7 because of the lack of $JHK$ measurements and inconsistencies between the extended and point source photometry.) }
\label{fig:SEDstars}
\end{figure}

\begin{figure}  
   \centering
 \label{fig:sedothers}
\includegraphics[width=\linewidth]{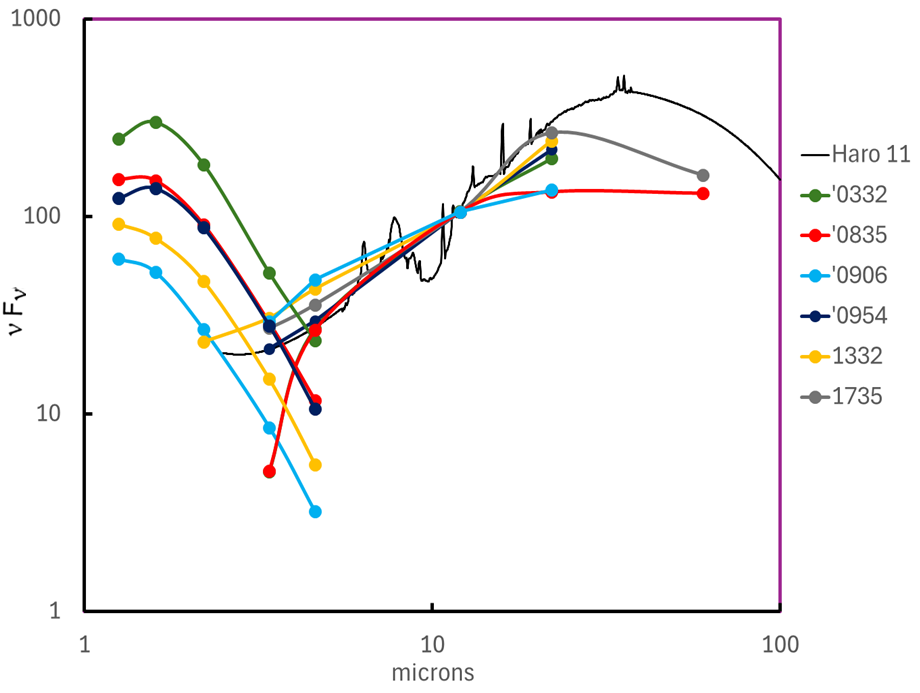}
    \caption{Similar to Figure~\ref{fig:SEDstars} but for the dwarf galaxies with emission line spectra indicating the presence of AGNs. Although the SED of Haro 11 is a good match for some, as anticipated in Section~\ref{sec:BPTAGNs}, three have excesses in the 2 - 5 $\mu$m range suggestive of a contribution by their AGNs (J090613.75+561015.5, J133245.62+263449.3, and J173501.25+570308.8) }
\end{figure}

  We show the SEDs of  11 potentially AGN-mimicking  galaxies in Figures ~6 and
  7, and compare them with the SED of Haro 11. All the spectra are normalized at
  WISE W3. The figures make a strong case that the Haro 11 SED is a maximal case
  for non-AGN galaxies; few if any non-AGN dwarf galaxies are likely to have
  ``hotter'' SEDs, i.e., ones with enhanced output in the 3 - 5 $\mu$m range.
 This suggests that the SED of Haro 11 provides a useful {\it ad hoc} template
 to use where there is a possibility for stellar-driven warm infrared excesses. 
 
 \citet[][Figure 1]{Derossi2018} and \citet{Remy2015} show how an increasing density of star formation results in  more dust heating and a ``hotter'' infrared SED. They also show that the hot SED of Haro 11  is due to its very dense star formation regions and that
 this template is a good fit to the galaxies with the highest density of star
 formation known at $z > 5$. It therefore provides a valid limiting hot template
 for typical galaxies out to z = 3 and beyond.

 \subsection{Identifying AGNs in more typical dwarf galaxies}

So far, we have focused on the dwarf galaxies with the hottest infrared SEDs.
These constitute no more than 0.2\% of a complete sample of local dwarf
galaxies. By adopting the Haro 11 SED as the fiducial template, one may
 exclude some galaxies containing AGNs to avoid a low level of
contamination by non-AGN cases. 

It may be possible to improve on this situation by observing the galaxies at
wavelengths $>$ 12 $\mu$m (rest). Figure~\ref{fig:seds} shows how this would
work. The SED of a typical AGN flattens past 10 $\mu$m and rolls over past about
25 $\mu$m \citep[e.g.,][]{Lyu2017}, while that of star forming galaxies
continues to rise to a peak at 60 - 150 $\mu$m \citep[e.g.,][]{Remy2015}. In Figure~\ref{fig:seds}, we compare the SEDs of five notorious heavily obscured AGNs with
that of Haro 11, and also show the WISE W4 fluxes for eleven dwarfs studied in
this paper. All the SEDs are normalized at $\sim$ 5 $\mu$m, i.e., at 4.6 $\mu$m
for W2 for the WISE photometry and at $\sim$ 5.2 $\mu$m for the obscured AGN
spectra. The figure shows a substantially greater relative 22 $\mu$m flux (W4)
for the star forming galaxies, with a difference already apparent for $\lambda
\ge$ 12 $\mu$m. Assuming that an obscured AGN dominates the SED\footnote{All
SED-fitting approaches operate under this assumption.}, this difference can be
exploited with photometry of relatively narrow bandwidth: WISE W3 is much too
broad, but the JWST/MIRI bands are suitable with $\Delta \lambda/\lambda \sim
0.2$. The MIRI 21 $\mu$m band allows use of this spectral discriminator out to z
$\sim$ 0.6. To exploit
this possibility, one also needs photometry to characterize the SED at shorter
wavelengths, to detect potentially confusing features such as silicate
absorption or PAH emission. The suite of MIRI photometric bands is well suited
to doing this. That is, the AGNs in dwarf galaxies can be identified
photometrically in a relatively complete fashion using survey data from MIRI
providing all the photometric bands, such as the CEERS \citep{Finkelstein2025}
or SMILES \citep{Rieke2024} surveys. For z $\gtrsim0.6$, we need future IR
missions such as PRIMA \citep{PRIMA2025} to extend the application of such selection techniques to
longer wavelengths \citep{Barchiesi2025}.

\begin{figure}[h] 
   \centering
\includegraphics[width=\linewidth]{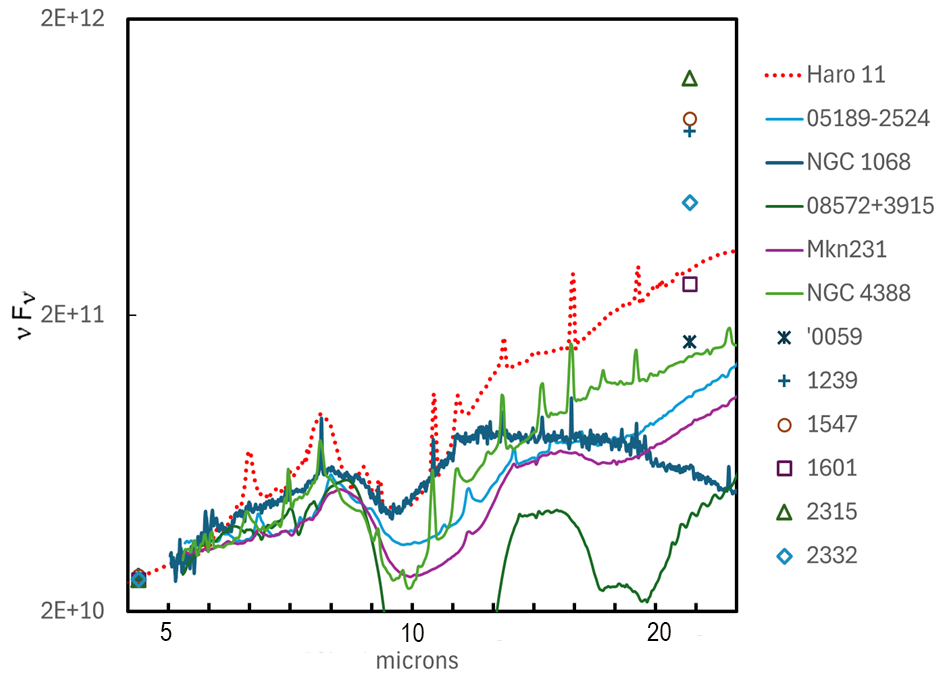}
    \caption{Comparison of Haro 11 mid-infrared spectrum with those of heavily obscured AGN examples. The figure also shows the WISE photometry of low metallicity  galaxies picked as possible AGN mimics. All the data are normalized at $\sim$ 5$ \mu$m. }
   \label{fig:seds}   
\end{figure}

There are, of course, many other methods to search typical dwarf galaxies for AGN that may escape spectroscopic detection. \citet{Messick2025} have carried out a large-scale search for variability, finding AGN candidates at a 2.4\% rate. \citet{Birchall2020} report a similar candidate identification rate from X-ray data. Neither of these approaches will reach any deeply embedded AGN that might be found exclusively in the infrared. However,   these studies emphasize the importance of multiple approaches to find complete AGN samples, as has been found also for more massive host galaxies. 

\section{Degeneracies}

 There is always a concern with photometric studies such as those discussed here that degeneracies could undermine the conclusions. We discuss this issue from high luminosities to low ones. 

\citet{Satyapal2018} discuss whether luminous star forming galaxies could be mistaken for AGN using WISE photometry. They conclude that this is a risk with two-band studies but the problem is largely mitigated if three bands are used. Together NIRCam and MIRI offer many bands over the WISE spectral range, all of which have been used in \citet{Lyu2024}, so we conclude that degeneracy in this range of luminosity  is not an issue, so long as a reasonably large suite of the bands offered by JWST is employed. 

In the high stellar mass case (log(M$_*$) $\ge$ 9.5), our  distinction between star formation and AGNs depends on two parameters: (1) the presence of PAH emission associated with star formation; and (2) the fitting of the infrared SED with a specific set of semi-empirical models for obscured AGN \citep{Lyu2022}.
The presence of PAH emission is virtually universal in star forming galaxies above a metallicity threshold that corresponds closely to the mass division between dwarf and massive galaxies \citep[e.g.,][]{Marble2010, Remy2015, Shivaei2024, Whitcomb2024}. In addition, the SED used for an obscured AGN is intrinsically flatter than the typical hot dust emission, so the obscured version has a distinct SED shape. In this paper we focused on a small subset of galaxies with extreme luminosity densities, which yields the hottest dust \citep{Remy2015}. A typical case would have lower temperature dust and a SED falling toward short wavelengths more steeply than the flat part of our assumed AGN SED. Between these two aspects of our modeling, degeneracy for the massive galaxy sample -- i.e., mistaking a star forming case for an AGN - should be very low, except at the higher redshifts (e.g., $z > 2.5$) where our spectral sampling of any infrared excess is limited. 

Degeneracy is a risk for dwarf galaxies where PAH emission can no longer be used as the signature of star formation. The conclusion we reach in this paper is that this issue can be avoided if the SED of Haro 11 is used as a limiting template and AGN candidates are only identified on the basis of SEDs with more short wavelength emission than this template.

\section{Conclusion}

This paper evaluates a few notable issues that may undermine the identification
of AGNs through infrared photometry. For relatively massive host galaxies,
$log(M) \ge 9.5$, where the mid-infrared excess powered by star formation should
be dominated by PAH features, we find that the major hazard is incorrect
redshifts. If the assigned redshift is too low, applying a SED template at that
redshift will predict excess emission at relatively short wavelengths. Since the
signature of an AGN is excess emission at $\lambda < 6 ~\mu$m, this can result
in false identifications. Otherwise, our high resolution rest-optical spectra
behaved as expected, showing star formation for host galaxies with heavily
obscured AGN  in most cases in rough agreement with the levels assigned
through SED fitting of photometry. 

We then considered the issue of dwarf galaxies with relatively ``hot'' infrared
excesses, which have been pointed out as mimicking AGNs \citep{Hainline2016,
Sturm2025}. As already pointed out by \citet{Hainline2016}, only a very small
fraction of dwarf galaxies have such SEDs. We show that the SED of Haro 11 can
be used as a limiting template; its SED is as ``hot'' as the most extreme star
forming dwarfs, so any galaxy with a hotter SED is likely to host an AGN (unless
it is of extremely low metallicity). However use of this template by itself may exclude some obscured AGNs with Haro 11-like (or cooler) SEDs. This issue can be mitigated by using photometry at $\sim $ 13 - 14
$\mu$m (rest), where true AGN SEDs are turning over while star forming ones are
rising toward the far infrared.

\section{Acknowledgements}

  Work on this paper was supported in part by grant 80NSSC18K0555, from NASA Goddard Space Flight Center to the University of Arizona. This work is based on observations made with the NASA/ESA/CSA James Webb Space Telescope. The data were obtained from the Mikulski Archive for Space Telescopes at the Space Telescope Science Institute, which is operated by the Association of Universities for Research in Astronomy, Inc., under NASA contract NAS 5-03127 for JWST. These observations are associated with program \#1207. The specific observations analyzed can be accessed via \url{https://archive.stsci.edu/hlsp/smiles} and \url{https://doi.org/10.17909/et3f-zd57}, doi = {10.17909/ET3F-ZD57}
    \url{http://archive.stsci.edu/doi/resolve/resolve.html}


\software {NumPy \citep{harris2020array}, Astropy \citep{Astropy2013, Astropy2018,Astropy2022}, Gelato \citep{Hviding2022}, IDL, Matplotlib \citep{Barrett2005},TOPCAT \citep{Taylor2005, Taylor2011}.}

\eject

\appendix

\appendix
\section{Analysis of spectra}
Here we present the expressions used to estimate the quantities and associated uncertainties reported in Tables 1 and 2.
For the H$\alpha$ Luminosity 
\begin{equation}
    Lum(H\alpha) = H\alpha_{obs}  \times 4\pi {DL(z)}^2 
\end{equation}
where  Lum(H$\alpha$) is the total luminosity emitted by the galaxy, H$\alpha_{obs}$ is the line flux obtained by fitting the observed spectrum assuming a Gaussian distribution, and the last term is the total area of a sphere centered on the galaxy, whose radius is the luminosity distance $DL(z)$ in Mpc derived from the observed redshift \citep{Hogg1999}  The associated uncertainty is estimated by assuming a 1\% error on the redshift ($\delta$$z$) and the flux measurement error:
\begin{equation}
    \Delta Lum(H\alpha) = \sqrt{(\frac {\delta H\alpha}{H\alpha})^2 + (\frac {\delta~z}{z})^2} \times Lum(H\alpha).
\end{equation}
The observed star formation rate follows the definition of \citet{Kennicutt2012}:
\begin{equation}
    SFR = Lum(H\alpha)/cx,
\end{equation}
where $cx$ is a constant such that log$_{10}$(cz) = 41.27, {\bf assuming a Kroupa IMF.}
The associated uncertainty is calculated as
\begin{equation}
\Delta SFR = \Delta Lum(H\alpha)/cx.
\end{equation}
The internal extinction correction ($A$) is derived using the  Balmer decrement H$\alpha$/H$\beta$, following \citet{Koyama2015}:
\begin{equation}
    A(H\alpha) = 6.53 \times log_{10}\frac {H\alpha}{H\beta} -2.98. 
\end{equation}
which has an estimated uncertainty of $\Delta~A\sim$ 0.24 magnitudes \citep[][equation 3]{Koyama2015}. This allows estimating the intrinsic luminosity of the source:
\begin{equation}
    L_{intrinsic} = L(H\alpha) * 10^{(0.4*A(H\alpha))}.
\end{equation}
The uncertainty is given by
\begin{equation}
\Delta L_{intrinsic} = \sqrt{(\Delta Lum(H\alpha)/Lum(H\alpha))^2 + (0.24/A(H\alpha))^2} \times L_{intrinsic}
\end{equation}
To estimate the star formation rate corrected for extinction SFR$_{corr}$ we use the intrinsic H$\alpha$ luminosity, e.g., \citet{Dominguez2013}:
\begin{equation}
    SFR_{corr} = L_{intrinsic}/cx,
\end{equation}
The uncertainty of the extinction-corrected star formation rate $\Delta$ SFR$_{corr}$ is calculated as:
\begin{equation}
     \Delta SFR_{corr} = (\Delta Lum(H\alpha) + Lum(H\alpha) \times 0.4 \times ln(10) \times \Delta H\alpha)/cx.
\end{equation}

\section{Display of spectra}

This appendix displays the NIRSpec MSA spectra used in the first part of the paper. 

\begin{figure*} 
   \centering
\includegraphics[width=0.8\textwidth]{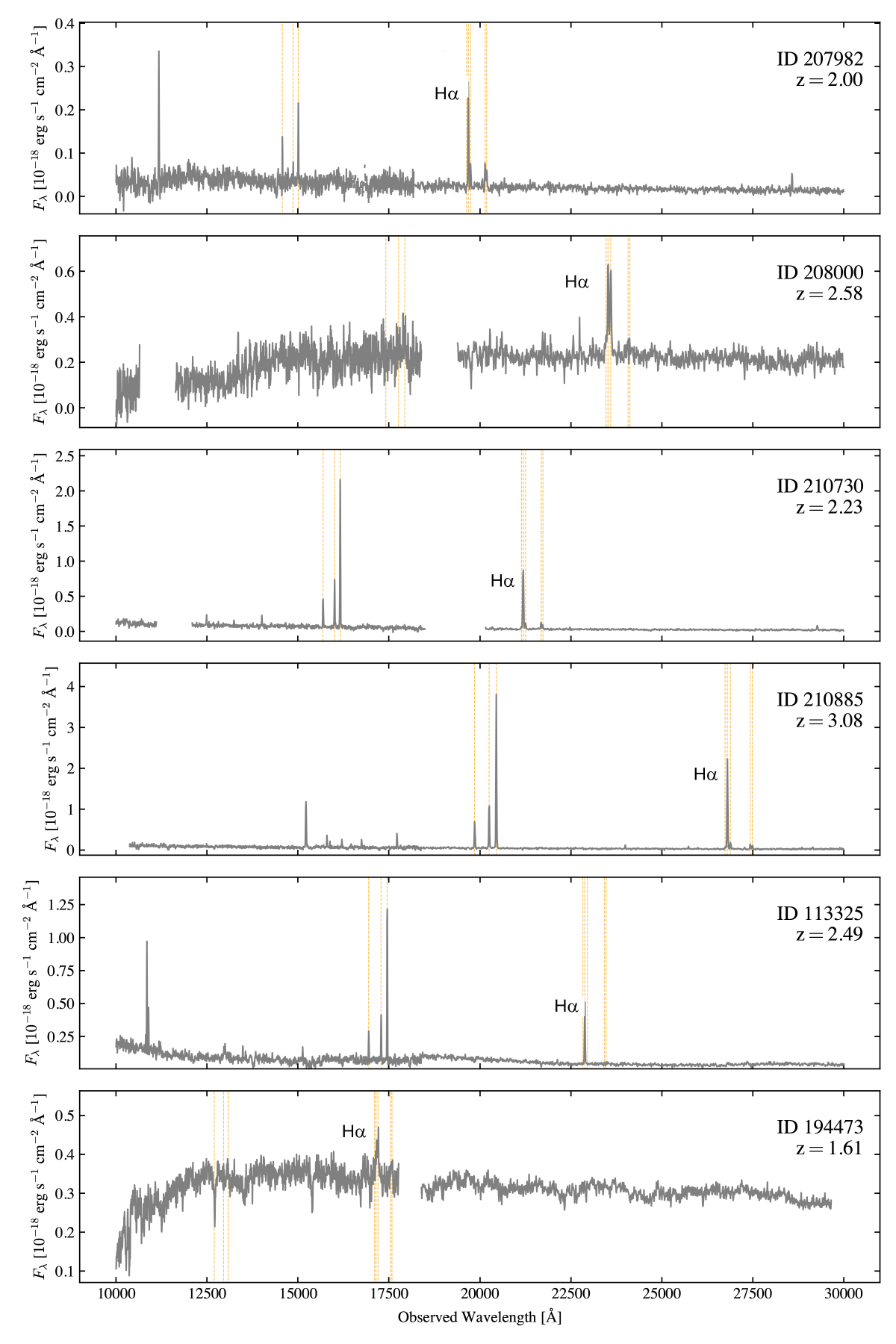}
    \caption{Spectra of six of the sources discussed in this paper. The H$\alpha$ lines have been indicated for quick orientation.}
    \label{fig:spectra1}
\end{figure*}

\begin{figure*} 
   \centering
   \figurenum{9}
\includegraphics[width=0.8\textwidth]{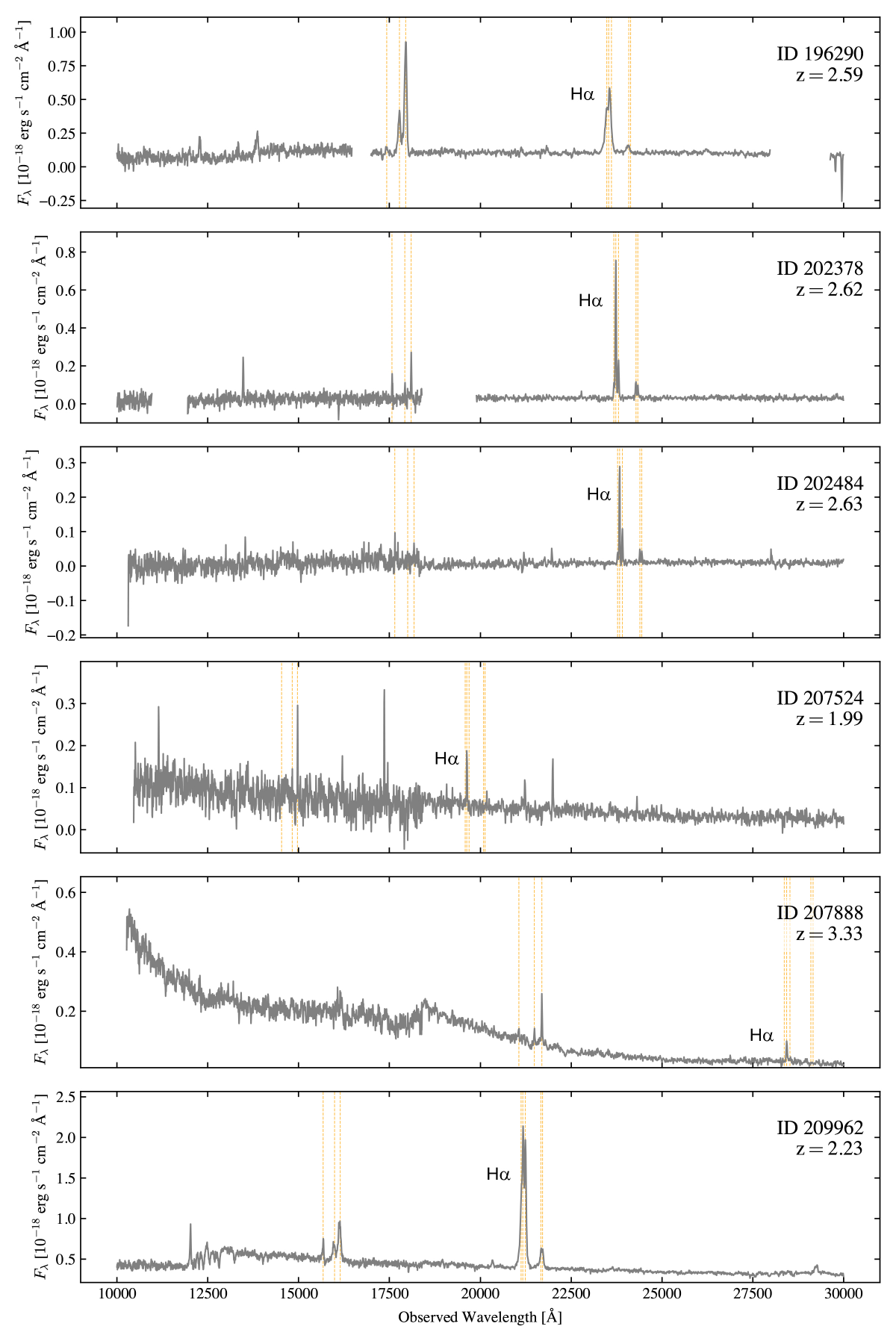}
    \caption{continued, spectra of six more of the sources discussed in this paper.}
    \label{fig:spectra1}
\end{figure*}

\begin{figure*} 
   \centering
   \figurenum{9}
\includegraphics[width=0.8\textwidth]{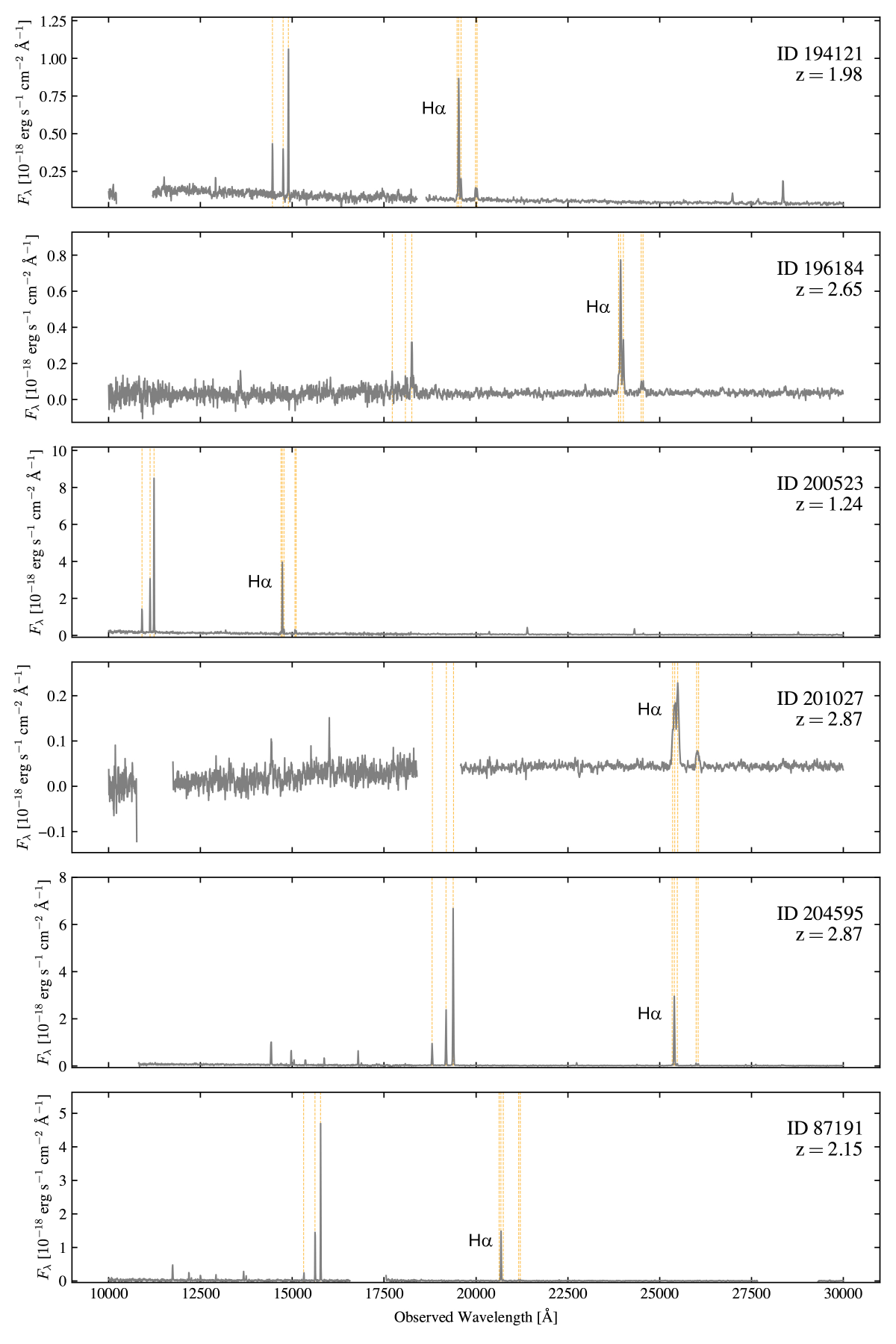}
    \caption{continued, spectra of six more of the sources discussed in this paper.}
    \label{fig:spectra1}
\end{figure*}

\end{document}